# Origins of Pressure-Enhanced Thermal Transport in Organic Semiconductors

Lukas Legenstein[1,2,*], Sandro Wieser[3], Michele Simoncelli[4], Egbert Zojer[2,†]

[1]Chair of Physics, Montanuniversität Leoben, Leoben, Austria
[2]Institute of Solid State Physics, Graz University of Technology, Graz, Austria
[3]Institute of Materials Chemistry, TU Wien, Wien, Austria
[4]Department of Applied Physics and Applied Mathematics, Columbia University, New York, USA

[*]lukas.legenstein@unileoben.ac.at, [†]egbert.zojer@tugraz.at

While pressure is known to dramatically alter the electronic properties of organic semiconductors, its impact on their thermal conductivity remains poorly understood. We combine machine learned potentials with the Wigner transport equation to compute the pressure-dependent thermal conductivity of crystalline naphthalene as a model system. When high-pressure reference data are included in the training, our simulations quantitatively reproduce the experimentally observed dramatic increase in thermal conductivity for compressed naphthalene. Most importantly, our results reveal the microscopic origin of this massive enhancement: pressure stiffens especially the intermolecular bonds, increasing the group velocities of heat-carrying phonons and simultaneously suppressing the scattering that impedes intraband (propagation) thermal transport. In contrast, interband (tunneling) transport is relatively weakened by a reduced spectral overlap between different phonon bands. These findings provide fundamental insights into heat conduction in soft molecular materials and suggest that strengthening intermolecular interactions, here, via applying pressure can be used to tune thermal transport in molecular crystals.

## Introduction

Understanding the pressure dependence of the lattice thermal conductivity of materials is a scientifically intriguing and at the same time challenging problem because, depending on the material, responses to compression can be significantly different.[1] What makes the response hard to predict is that it can range from monotonic increases or decreases to an anomalous, non-monotonic pressure dependence marked by an initial increase of the thermal conductivity followed by a subsequent decrease.[2–4] As a further complication, measuring lattice thermal conductivity at elevated pressures is anything but straightforward and requires, e.g., specialized characterization techniques integrated into megabar diamond anvil cells.[1,5–9] Because of such technical challenges, simulations have become indispensable tools that provide fundamental and complementary insights.

For example, modern molecular dynamics and lattice dynamics approaches enable the simulation of heat transport at pressures that would be experimentally inaccessible,[2,10] they allow distinguishing competing microscopic mechanisms,[3,11,12] and they make it possible to track structure-to-property relationships in a controlled and systematic way.[13–15] Moreover, simulations allow the observation of pressure-dependent effects at a level of detail that is difficult to achieve in experiments.

So far, the bulk of available pressure-dependent studies has focused on comparably simple, covalently bonded crystals,[1] such as silicon,[16] binary compound semiconductors,[3] materials of the Earth's core and mantle like MgO,[17] and various chalcogenide thermoelectric materials, like PbTe and PbSe.[13] In fact, to our knowledge, no computational studies have examined pressure-dependent heat transport in organic materials. The only related material that has been studied is $C_{60}$, which is strictly speaking not organic but still often classified as an n-type organic semiconductor.[18] The relative pressure enhancement of its thermal conductivity, $\frac{1}{\kappa_0}\frac{\Delta\kappa}{\Delta p}$, is exceptionally large, increasing from 135% per GPa (between 0 GPa and 5 GPa) to 200% per GPa (between 5 and 20 GPa). This by far exceeds the enhancements observed for most inorganic materials.[1]

The exceptional behavior of $C_{60}$ calls for the study of related van der Waals-bonded materials, like actual aromatic hydrocarbons. This is relevant, as it also provides fundamental insights useful for the continuously growing field of organic electronics.[19–22] Organic semiconductors (OSCs) are notoriously soft and compressible because of weak non-covalent interactions between the molecules.[23] Moreover, even moderate pressures, applied intentionally during processing or unintentionally during operation, can change the molecular arrangement and, thus, intermolecular coupling and charge transport properties.[24] In fact, the charge carrier mobility of organic semiconductors has been shown to significantly improve under pressure, as demonstrated for several molecular semiconductors including naphthalene.[25,26] This motivates the question of how the thermal conductivity of organic semiconductors is affected by high pressure.

Our focus in the present study lies on crystalline naphthalene, the simplest member of the acene series. A particular advantage of naphthalene is that for this material, experiments by Ross et al. show a similarly strong pressure enhancement of the thermal conductivity as in $C_{60}$.[27] Thus, any simulation can be compared to high-quality experiments before it is used to analyze the underlying mechanisms. Moreover, the crystal structure and the vibrational properties of naphthalene have been extensively characterized over a wide range of pressures.[28–33] On more technical grounds, the advantages of naphthalene are the crystal's manageable unit cell size and that, at least according to X-ray diffraction studies, it does not undergo any structural phase transitions within the considered temperature and pressure ranges.[34,35,32,36] Like all members of the acene family, naphthalene crystallizes in the so-called herringbone structure, a type of packing common to the majority of π-conjugated molecular crystals and in particular the series of unsubstituted acenes.[37–40] This makes naphthalene—despite its large fundamental gap, which renders it effectively an electrical insulator—an instructive reference system for multiple types of OSCs.

In our recent work, we showed that to accurately describe thermal conductivity in OSCs at ambient pressure, it is critical to go beyond the semiclassical Peierls-Boltzmann equation[41], describing particle-like transport via phonon propagation, by considering inter-mode coupling (or "phonon tunneling").[42] Both microscopic heat transport mechanisms can be comprehensively described using the Wigner transport equation,[43,44] which formally unifies and extends the known propagative and mode-coupling theories for heat transport in crystals and amorphous solids. By operating in reciprocal space, it also allows disentangling the contributions of different types of phonons prevalent in the involved phonon band structures of organic semiconductors[45]. How the resulting highly complex phonon transport phenomena are influenced by pressure in OSCs is, however, largely unexplored.

The technical challenge encountered here lies in accurately modeling pressure-dependent changes in atomic vibrations while maintaining computational efficiency to simulate hundreds of thousands of displaced supercells for obtaining anharmonic phonon scattering rates. We address this challenge by applying an active-learning[46,47] workflow[42,48,49] to train specialized moment tensor potentials[50] (MTPs) that achieve close to *ab initio* accuracy over a wide pressure range (>10 GPa). Such machine-learned potentials have been shown to produce quantitatively accurate thermal conductivities for several classes of complex materials.[42,48,51–54]

Using highly accurate MTPs, we reproduce the experimentally observed pressure-induced compression of the crystal lattice[32] as well as lattice parameters calculated with density functional theory. We directly solve the Wigner transport equation for structures under varying pressure. Crucially, our results show excellent agreement with experimental thermal conductivities at both ambient and elevated pressures. This successful validation of our approach enables us to provide a detailed and reliable discussion of how the pressure dependence of the thermal conductivity can be traced to modifications in the phonon band structure, group velocities, and phonon lifetimes, eventually suggesting a design strategy for enhancing heat transport in OSCs.

# Results

## Naphthalene's crystal structure under pressure

One of the characteristics of naphthalene that is particularly well-studied is its crystal structure (illustrated in Figure 1a) as a function of temperature and pressure.[32,35,55–57] In previous studies, it has already been established that naphthalene's low-temperature crystal structure can be excellently reproduced by a high-accuracy moment tensor potential (MTP)[42] or by the underlying dispersion-corrected density functional theory method (PBE[58] with Grimme's D3BJ[59] correction)[60]. Accurately reproducing the pressure-induced structural changes is an important prerequisite for lattice dynamics simulations of heat-transport phenomena. Therefore, we first discuss the changes of naphthalene's lattice parameters under elevated hydrostatic pressure before describing how they are related to changes of the material's thermal conductivity.

X-ray diffraction experiments under varying pressure, such as those by Likhacheva et al.,[32] reveal that naphthalene undergoes significant, anisotropic, and non-linear compression. Figure 1b (black crosses) shows the experimentally observed pressure-induced changes in lattice parameters *a*, *b*, and *c* as (negative) strain. In the examined pressure range, compression primarily reduces intermolecular distances rather than intramolecular bond lengths.[32] An important initial question is whether a basic MTP (trained on uncompressed naphthalene structures in ref. 42) would be capable of accurately describing the material also at higher pressure. As illustrated in the comparison between experimentally measured lattice parameters and those obtained from strained structure relaxations using the said MTP in Figure 1b (open squares connected by dashed lines), this is not the case. Although the calculated pressure-induced strain of the lattice parameters *b* and *c* follows the experimental trend, the strain predicted for parameter *a* shows an erroneous, linear decrease with pressure and a significantly overestimated compression above 1.5 GPa. Even worse, beyond 2.5 GPa, the structure relaxations using the basic MTP even fail completely (i.e., the relaxation in MLIP-2[61,62] stops to avoid atoms getting too close).

To resolve this problem, it is necessary to understand how the crystal structure responds to pressure. Acene molecules crystallize in a herringbone arrangement with angles between the molecular planes distinctly below 90° (Figure 1c). Under pressure, the molecules tilt, and the angle becomes more acute. Not surprisingly, the basic MTP cannot correctly capture such comparably complex changes in molecular arrangements. Notably, the problem is not intrinsic to the *ab initio* technique used in the MTP parametrization, as geometry optimizations with the PBE exchange functional[58] including a Becke-Johnson damped D3 van-der-Waals correction[59] (black, open spheres) excellently reproduce the experiments. This calls for the explicit inclusion of compressed reference structures in the MTP training procedure.

The latter relies on molecular dynamics simulations with active learning in VASP,[46,47] which generates training datasets comprising diverse *ab initio*-calculated reference configurations (for details, see Methods section). The original "basic" training dataset consists of 459 uncompressed structures of 72 atoms (four molecules) each. To generate "multi-pressure" MTPs, we augment the training data with an additional 553 *ab initio* configurations, obtained in active-learning simulations performed at external pressures of 0.5 GPa, 2 GPa, and 10 GPa (see Methods). As shown in Figure 1b and 1d, the multi-pressure MTP (solid lines) successfully reproduces the experimentally observed non-linear and anisotropic compression of naphthalene, including the change in the herringbone angle. In passing, we note that the multi-pressure model remains stable across an even wider pressure range than displayed and can be used for structural relaxation up to 15 GPa (see Figure S2). At even higher pressures, the multi-pressure MTP also fails, presumably again because the compression too strongly deviates from the training conditions.

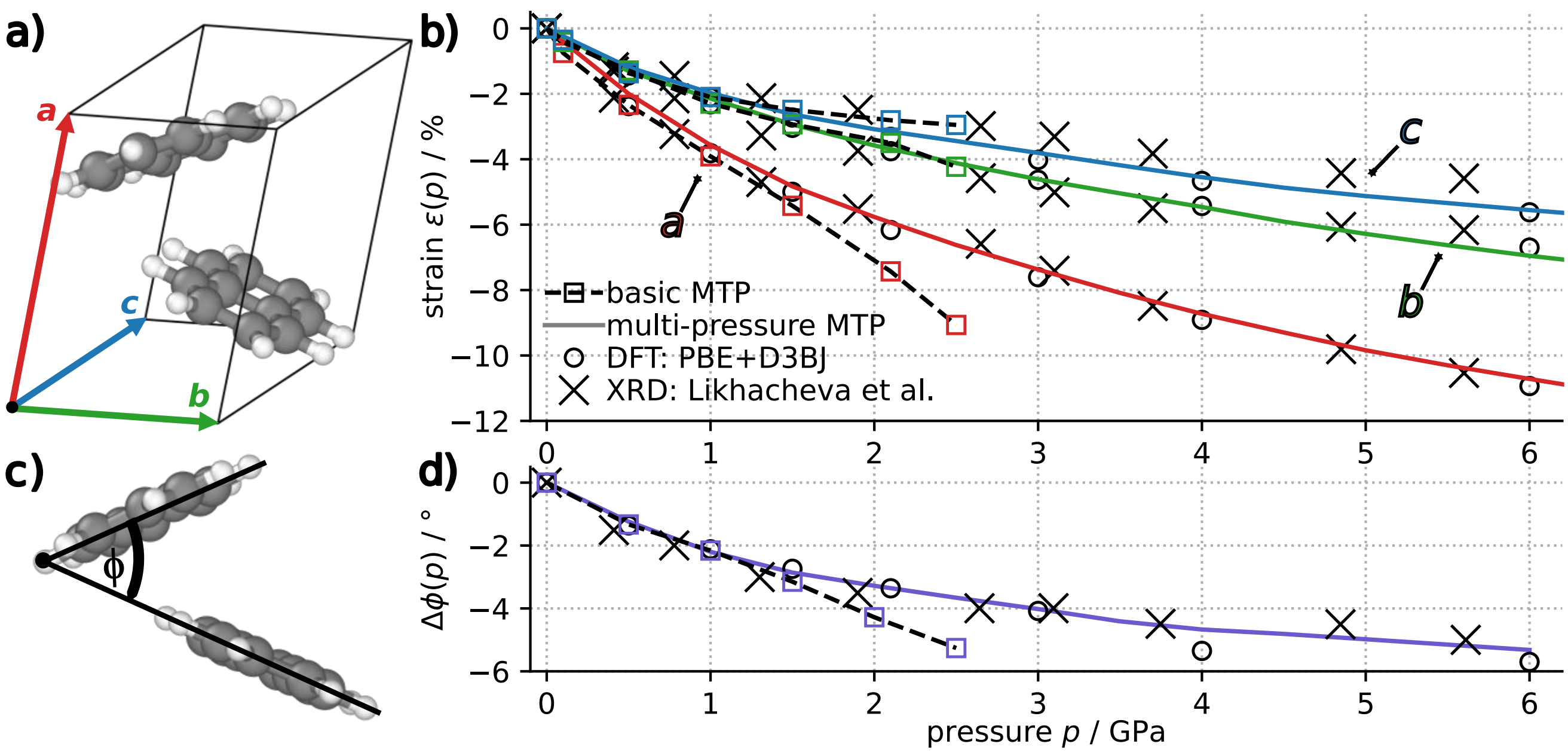

**Figure 1:** *Evolution of naphthalene's crystal structure under hydrostatic pressure: (a) naphthalene's crystal structure, illustrated with Ovito*[63]*; (b) strain, i.e., relative change, of lattice parameters a (red), b (green), and c (blue) as a function of pressure; (c) illustration of the intermolecular herringbone angle; (d) change of the intermolecular herringbone angle as a function of pressure. Data from X-ray diffraction experiments by Likhacheva et al.*[32] *are presented by black crosses. Theory results from enthalpy optimizations with DFT are presented by black circles, from those with the basic MTP by open squares and dashed lines, and from those with the multi-pressure MTP by solid lines.*

Note that Figure 1 shows relative changes in lattice constants instead of absolute values to better illustrate the effects of pressure. The absolute values are additionally influenced by thermal expansion, which is not included in the gradient-based unit-cell optimizations performed with DFT and the two MTPs. While the impact of naphthalene's thermal expansion on its thermal conductivity is not entirely negligible,[64] it is smaller than pressure-driven changes and does not alter the observed pressure-dependent trends. A direct comparison of the actual lattice parameters rather than the strains can be found in Figure S2 and the impact of thermal expansion on heat transport is discussed in more detail in section 2 of the Supporting Information.

**Pressure-enhanced heat transport and changes to its microscopic mechanisms**

As a next step, we apply the aforementioned MTPs to compute the interatomic force constants for crystal structures under varying pressures. From those force constants, we obtain pressure-dependent phonon band structures, group velocities, and three-phonon scattering rates. Details on these simulations can be found in the Methods section. The aforementioned properties are then used to compute the lattice thermal conductivity, $\kappa$, by solving the Wigner transport equation[43,44] in the relaxation-time approximation—a strategy that has been shown to accurately model thermal transport in OSCs crystals.[42] Within this approach, the total thermal conductivity can be written as the sum of two terms:

$$\kappa_{\mathrm{tot}} = \kappa_{\mathrm{P}} + \kappa_{\mathrm{C}}. \quad (1)$$

The first term, $\kappa_{\mathrm{P}}$, accounts for the diagonal elements ($s=s'$) of the group velocity operator $V(\boldsymbol{q})$ for phonon bands $s$ and $s'$. It is equivalent to the Peierls-Boltzmann conductivity expression that describes thermal transport via intraband phonon propagation and is given by:[41]

$$\kappa_{\mathrm{P}}^{\alpha\beta} = \frac{1}{\Omega N} \sum_{\boldsymbol{q},s=s'} C(\boldsymbol{q})_s V^{\alpha}(\boldsymbol{q})_{s,s} V^{\beta}(\boldsymbol{q})_{s,s} \frac{1}{\Gamma(\boldsymbol{q})_s}. \quad (2)$$

The second term, $\kappa_{\mathrm{C}}$., describes the coherence thermal conductivity that results from interband coupling, often referred to as "phonon tunneling", between different bands ($s \neq s'$). It can be written as:

$$\kappa_{\mathrm{C}}^{\alpha\beta} = \frac{1}{\Omega N} \sum_{\boldsymbol{q},s\neq s'} \frac{\omega(\boldsymbol{q})_s + \omega(\boldsymbol{q})_{s'}}{4} \left[\frac{C(\boldsymbol{q})_s}{\omega(\boldsymbol{q})_s} + \frac{C(\boldsymbol{q})_{s'}}{\omega(\boldsymbol{q})_{s'}}\right] \quad (3)$$
$$\times V^{\alpha}(\boldsymbol{q})_{s,s'} V^{\beta}(\boldsymbol{q})_{s',s} \frac{\frac{1}{2}[\Gamma(\mathbf{q})_s + \Gamma(\mathbf{q})_{s'}]}{[\omega(\boldsymbol{q})_s - \omega(\boldsymbol{q})_{s'}]^2 + \frac{1}{4}[\Gamma(\boldsymbol{q})_s + \Gamma(\boldsymbol{q})_{s'}]^2}.$$

In the above equations, $\Omega$ denotes the volume of the primitive cell and $N$ represents the number of sampled wavevectors $\boldsymbol{q}$ within the Brillouin zone. The quantities $\omega$, $C$, and $\Gamma$ correspond to the phonon frequencies, mode heat capacities, and phonon scattering rates resulting in finite linewidths (and lifetimes). The Cartesian components of the tensors are labeled by the indices $\alpha$ and $\beta$.

When comparing theory and experiments performed on compacted naphthalene powder, we consider the total orientation-averaged thermal conductivity, $\kappa^{\mathrm{av}}$, given by the trace of the thermal conductivity tensor:

$$\kappa^{\mathrm{av}} = \frac{1}{3}(\kappa^{xx} + \kappa^{yy} + \kappa^{zz}). \quad (4)$$

Although this arithmetic mean represents only the upper-bound estimate of the thermal conductivity for a polycrystalline sample,[65] our previous work[42] on naphthalene showed that the difference between the upper- and lower-bound estimates (the harmonic mean) is negligibly small.

Figure 2 compares the simulated and measured temperature- and pressure-dependent values of $\kappa^{\mathrm{av}}$ of naphthalene. We note in passing that the temperature-dependent calculations account for the dominant changes in phonon occupation and resulting changes in phonon scattering rates while the comparatively minor effect of thermal expansion is not considered (see also discussion in [42] and in section 2 in the Supporting Information). For uncompressed naphthalene (0 GPa), the lattice thermal conductivity predicted with the basic MTP (blue, open squares) is in excellent agreement with experiments by Überreiter and Orthmann[66] (green triangles), as previously discussed.[42] When increasing the pressure to 2.1 GPa, the thermal conductivity increases significantly at all temperatures, as shown by the experiments by Ross and colleagues[27] (drawn as a blue, filled curve). Qualitatively, this effect is reproduced by the basic MTP, but quantitatively, the basic MTP systematically underestimates the conductivity at 2.1 GPa by up to 67% compared to experiments in the presented temperature range.

This underestimation is insofar unexpected, as the basic MTP overestimates the strain of the lattice at 2.1 GPa (see Figure 1b), which would artificially raise group velocities and, consequently, lead to an overestimation of the predicted thermal conductivity. The fact that the opposite occurs demonstrates that the situation is more complex and that the MLP trained solely under ambient conditions is unsuited for simulations at elevated pressure. This assessment is supported by the data for the pressure dependence of the thermal conductivity in Figure 2b, where simulations at 300 K (squares) are compared to the experiments by Ross et al.[27] at 297 K (crosses).

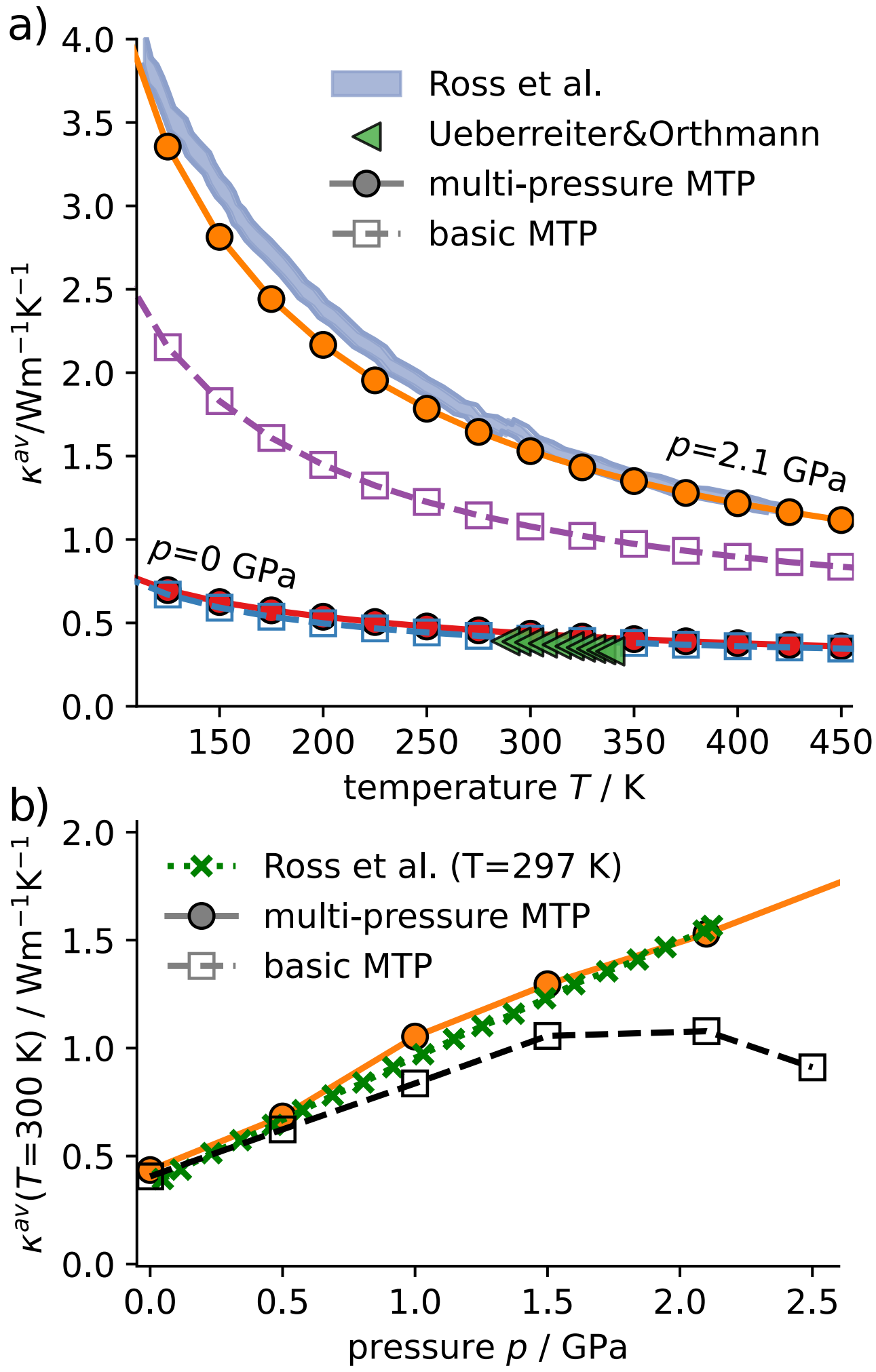

**Figure 2:** *Lattice thermal conductivity of naphthalene under pressure: (a) Temperature dependence of the orientation-averaged conductivity at 0 and 2.1 GPa, comparing predictions from the basic (open squares) and multi-pressure (filled circles) MTPs with experimental data from Ross et al.[27] (blue line at 2.1 GPa) and Überreiter & Orthmann[66] (green triangles at 0 GPa). (b) Pressure dependence of the orientation-averaged thermal conductivity at 300 K, with MTP predictions compared to experimental data from Ross et al.[27] at 297 K (green crosses). Theoretical data up to 15 GPa is shown in Figure S7.*

The situation is fundamentally different for the multi-pressure MTP trained on the augmented set of reference structures, comprising also high-pressure configurations (see above). This MTP provides a quantitatively highly accurate description of the temperature-dependent thermal conductivity at 2.1 GPa (orange circles in Figure 2a). It also reproduces the linear increase of the thermal conductivity with external pressure seen in experiments (c.f., Figure 2b) with essentially identical rates of increase of $\kappa^{av}$ in the experiments (0.56 $Wm^{-1}K^{-1}GPa^{-1}$) and in the simulations (0.54 $Wm^{-1}K^{-1}GPa^{-1}$). Beyond the experimentally considered pressure range of up to 2.1 GPa, the multi-pressure MTP predicts a further linear increase of $\kappa^{av}$ up to 15 GPa (see Figure S7). Importantly, the quality of the description for the 0 GPa structure of naphthalene (red circles in Figure 2a) does not deteriorate when using multi-pressure MTPs. This shows that augmenting the training data by high-pressure configurations creates an MLP that works reliably across the entire investigated pressure range, making it a general-purpose model for the systems at hand.

With the accuracy of the multi-pressure MTP demonstrated, it is next used to examine the underlying transport mechanisms. First, the relative impact of phonon tunneling (Equation 3) is assessed relative to phonon propagation within the Peierls-Boltzmann framework (Equation 2). This is relevant, considering that at ambient pressure the inclusion of phonon tunneling is crucial for accurately describing the thermal conductivity of acenes.[42] Figure 3a illustrates the pressure-dependent propagation ($\kappa_P$) and tunneling ($\kappa_C$) thermal conductivities relative to their 0 GPa values for different temperatures. This is done using a log-scale to accentuate the different rates of change under compression. Both, $\kappa_P$ and $\kappa_C$ increase with pressure (with the exception of $\kappa_C$ at 100 K). The pressure-induced enhancement of $\kappa_P$ is, however, roughly one order of magnitude larger.

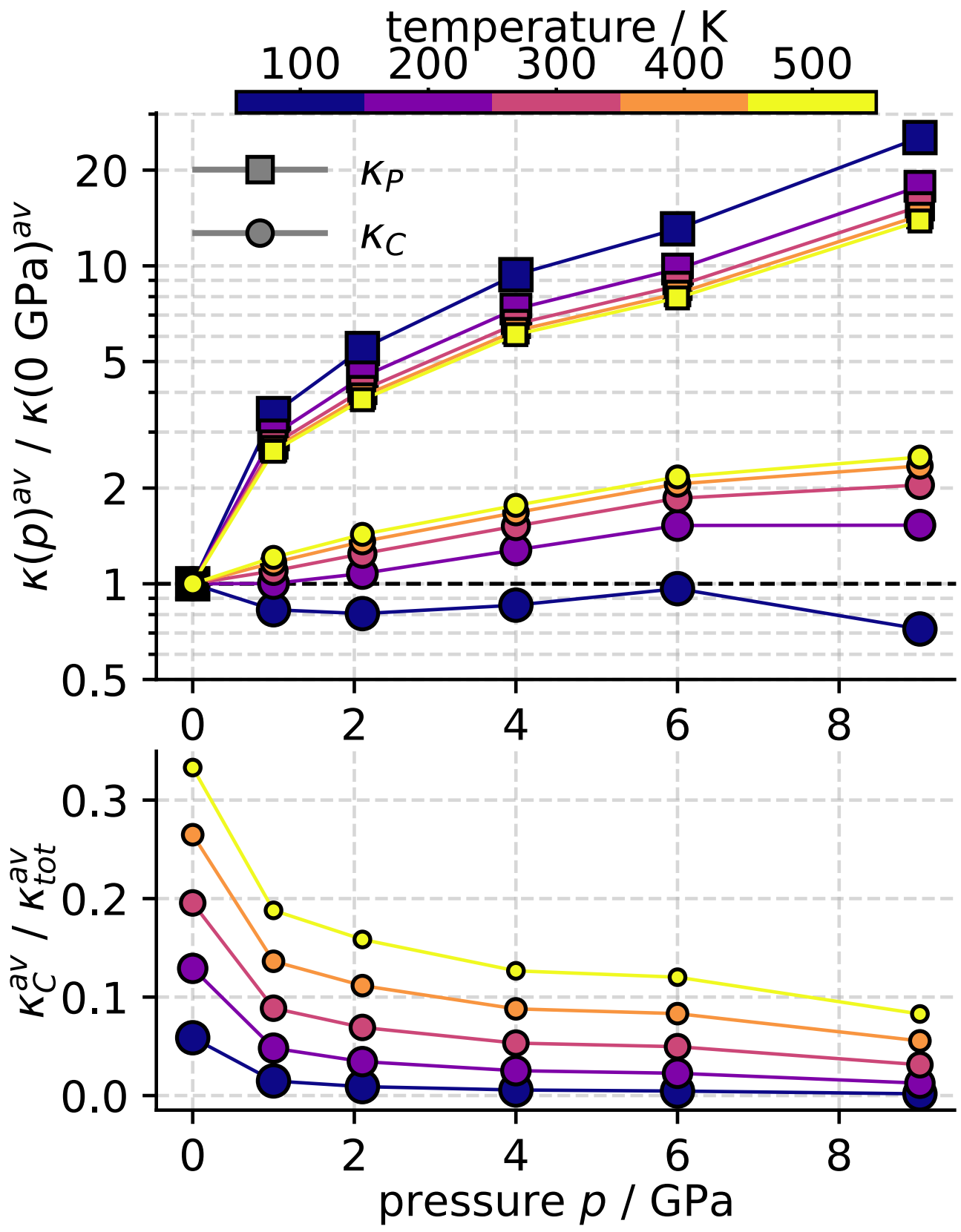

**Figure 3:** *Pressure-induced change of the orientation-averaged propagation, $\kappa_P$, and interband tunneling thermal conductivity, $\kappa_C$: (a) propagation (squares) and tunneling (circles) thermal conductivity as a function of pressure relative to the values of the uncompressed structure (0 GPa) calculated for different temperatures (see color scale on top of the plot). (b) Tunneling conductivity relative to total conductivity as a function of pressure.*

For instance, at 300 K, $\kappa_P$ rises dramatically from 0.35 $Wm^{-1}K^{-1}$ at 0 GPa to 5.39 $Wm^{-1}K^{-1}$ at 9 GPa, whereas $\kappa_C$ only doubles from 0.09 $Wm^{-1}K^{-1}$ to 0.17 $Wm^{-1}K^{-1}$. The diminishing relative contribution of $\kappa_C$ at elevated pressures is further illustrated in Figure 3b by the ratio between $\kappa_C$ and $\kappa_{tot}$. The effect is particularly dramatic at low temperatures, where at 100 K and beyond 2.1 GPa the tunneling contribution becomes essentially negligible. Overall, the results in Figure 3b clearly show that for naphthalene, the inter-band coupling transport mechanism progressively loses relevance with increasing pressure. Thus, the following discussion primarily focuses on pressure-induced changes of $\kappa_P$.

**Changing roles of acoustic and optical phonons under pressure**

To understand the strong pressure-induced enhancement of the propagation conductivity it is necessary to identify the nature of the phonons primarily responsible for this transport mechanism. A key distinction is to be made between acoustic and optical phonons. Acoustic phonons are associated with large displacements of molecules as rigid units and are widely regarded as the primary heat carriers in most semiconductors.[67] The associated vibrational branches typically form the three lowest-frequency phonon bands that usually show the strongest dispersion. However, as shown in Figure 4a, the phonon band structure of OSCs can be quite intricate, with both acoustic and multiple optical bands present in the low-frequency region. This complexity leads to numerous (avoided) band crossings, which can be accompanied by mode hybridization. Notably, the complexity of the band structures and the number of optical bands within this frequency region increases when the molecules in the crystal become larger.[68] Consequently, in OSCs and other complex crystals, distinguishing acoustic from optical phonons is not necessarily a trivial task that can be accurately achieved by a mere “visual” inspection of the band-structure. Therefore, it is useful to quantitatively categorize the phonon modes based on the nature of the associated atomic displacements encoded in their polarization vectors. This is achieved by calculating the acoustic participation ratio ($APR$)[64,69] following the procedure described by Kamencek et al.[69]. As is common for participation ratios, the $APR$ measures the extent to which all atoms participate in a vibrational mode, but it goes a step further by also factoring in the uniformity of the displacements' spatial alignments. The $APR$ of a specific wave vector $q$ in band $s$ is defined as

$$APR(\boldsymbol{q})_s = \frac{2}{N(N+1)} \frac{\left|\sum_{i=1}^{3N}\sum_{j\geq i}^{3N} \frac{(e(\boldsymbol{q})_s^i)^\dagger e(\boldsymbol{q})_s^j}{\sqrt{m_i m_j}}\right|^2}{\sum_{i=1}^{3N}\sum_{j\geq i}^{3N} \left|\frac{(e(\boldsymbol{q})_s^i)^\dagger e(\boldsymbol{q})_s^j}{\sqrt{m_i m_j}}\right|^2} \quad (5).$$

It is calculated from the inner products of eigenvectors $e(\boldsymbol{q})_s^i$ and $e(\boldsymbol{q})_s^j$ of atoms $i$ and $j$, weighted by their respective masses $m$, with $N$ being the number of atoms in the unit cell. By definition, acoustic phonons typically have an $APR$ close to 1, while optical phonons have an $APR$ close to 0.

A visual presentation of the $APR$ is shown for naphthalene's band structure at 9 GPa via a color scale in Figure 4a, with purely optical bands displayed in blue and purely acoustic bands in red. Intermediate colors (orange to green) directly reveal phonons with a hybrid character, e.g., towards the Brillouin zone boundaries as well as along the X—A path where the acoustic and the optical translation branches have degenerate energies. It also visualizes avoided band crossings by abrupt and localized changes in color between two adjacent bands, which is indicative of the transfer of their polarization vectors.[68]

An additional distinction between longitudinal acoustic (LA) and transverse acoustic (TA) branches can be made by quantifying the spatial alignment between atomic displacements of a mode and the respective wave vector $\boldsymbol{q}$. The mathematical details of the procedure used to determine the "longitudinality" of the modes are provided [67], while a visual presentation of the property is provided in section 5 (Figure S8) of the Supporting Information.

Next, we use the *APR* and the "longitudinality" to decompose the thermal conductivity described in Equation 2 into contributions from LA, TA, and optical phonons. First, the acoustic contribution is determined by weighting the mode-resolved conductivities with the *APR*, while the optical contribution is weighted by (1-*APR*), i.e., it is the residual to $\kappa_P$. We then further separate the acoustic contribution by applying the "longitudinality" as an additional weight to isolate the LA phonon conductivity, where now the TA contribution is the residual.

The resulting pressure-dependent ratios of LA (plus symbols), TA (crosses), and optical phonon contributions (circles) to the propagation conductivity are illustrated in Figure 4b for 100 K, 200 K, 300 K, 400 K, and 500 K. They reveal several interesting aspects: at ambient pressure and temperatures above 100 K, optical phonons contribute more than 50% to the (propagation) thermal conductivity. This challenges conventional wisdom that acoustic phonons are the primary heat carriers. However, as pressure increases to 2.1 GPa, the share of optical phonons decreases sharply, beyond which the decline becomes more gradual and roughly linear. At 9 GPa, optical phonons account for approximately 25% of the particle-like conductivity for low temperatures and ca. 35% at 300 K and above.

Although LA bands have the largest individual mode contributions, collectively they carry less heat than TA and optical modes across all temperatures and pressures. Overall, the pressure-induced increase of the relative contribution of the LA modes is less pronounced than that of the TA modes, and essentially saturates around 25% above 2.1 GPa. In contrast, under compression the TA modes become dominant, generating between 40% and 50% of the propagation conductivity at 9 GPa. A similar analysis for $\kappa_C$ and the decomposed absolute values of $\kappa_P$ are presented in Figure S9 and Figure S10.

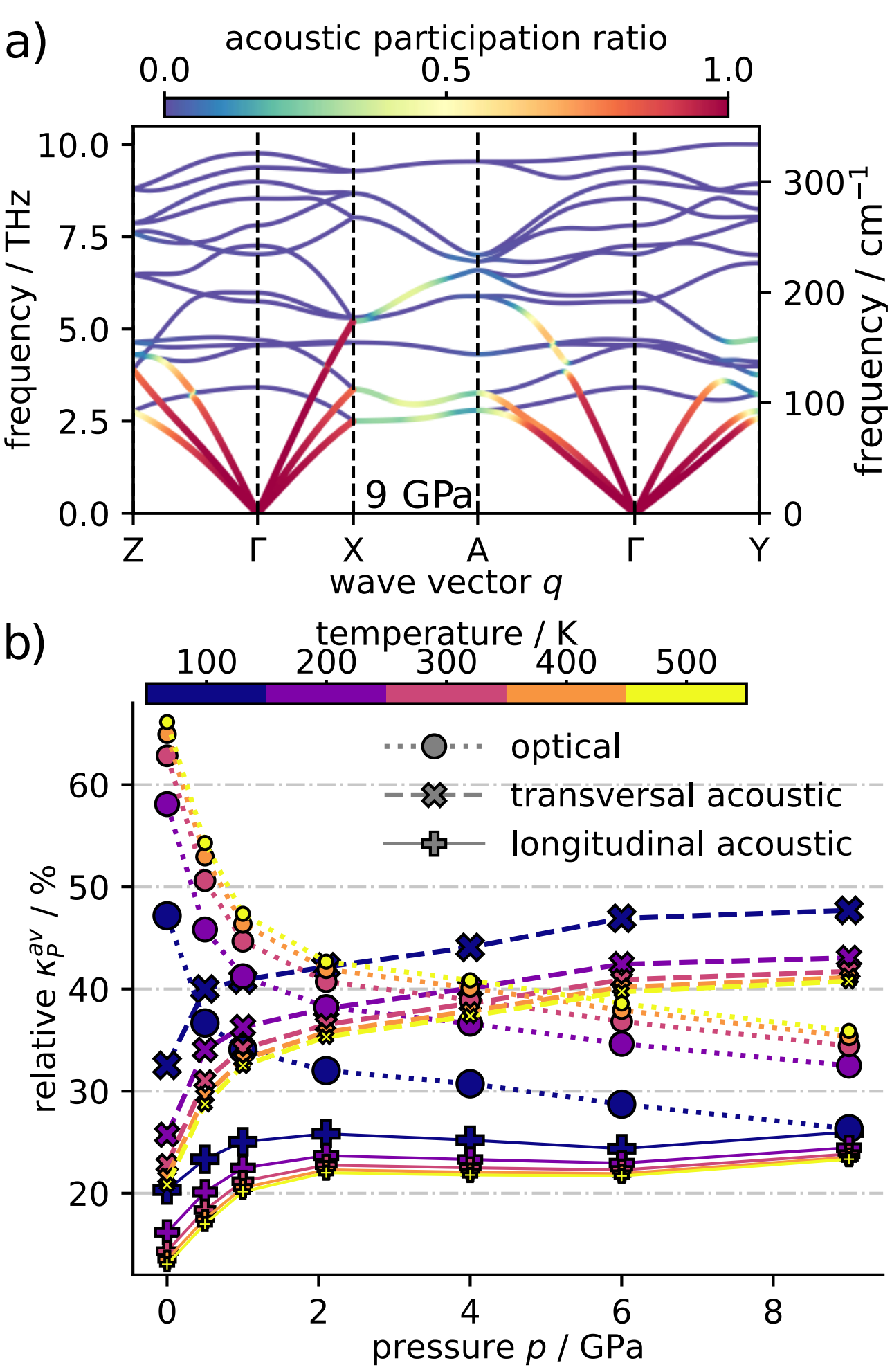

**Figure 4:** *Decomposition of thermal transport based on phonon characters: (a) Phonon band structure of naphthalene at 9 GPa of hydrostatic pressure. The color code indicates the acoustic participation ratio of the individual phonon modes (see color bar), where those in red (blue) can be characterized as acoustic (optical) phonons. Intermediate colors (orange to green) indicate modes of mixed character. (b) Relative contributions of optical (circles), transverse acoustic (crosses), and longitudinal acoustic (plus symbols) phonons to the orientation-averaged propagation thermal conductivity as a function of pressure. The color of the data points denotes the temperature according to the color bar at the top of panel b.*

**Physical origins of naphthalene's exceptional pressure-enhanced thermal transport**

To understand these trends, it is necessary to consider the role of individual phonon bands. An important peculiarity of molecular crystals in this context is that they exhibit two distinct types of vibrational branches with a detailed discussion of the nature of all low-lying bands of the acene crystals provided in [45]. In short, the first type of vibrations — often characterized as intermolecular or external — involves the rigid-body motion of entire molecules through translational and rotational displacements. These phonon branches with the lowest frequencies are governed by weak non-covalent intermolecular interactions and the relatively large molecular masses. The acoustic bands and the nine lowest-lying optical bands belong to this group. The second type comprises intramolecular (or internal) vibrations, which either distort the molecular backbones or involve bond stretching and bending in specific groups of atoms. The corresponding 96 branches constitute the majority of naphthalene's vibrations and cover an exceptionally broad frequency spectrum extending up to around 94 THz (3140 cm$^{-1}$), where vibrations associated with the C-H bonds are found. In the absence of externally applied pressure (i.e., at 0 GPa) all intermolecular bands are below ~4 THz (135 cm$^{-1}$), as shown in the phonon band structure and in the densities of states in Figure 5a and d. At ambient pressure, they remain well-separated from the lowest internal vibrations by a band gap between 4.1 THz (137 cm$^{-1}$) and 5.5 THz (183 cm$^{-1}$). This gap is illustrated by the orange rectangle in Figure 5a.

Not unexpectedly, hydrostatic pressure affects external and internal vibrations differently. Compression primarily reduces intermolecular distances,[32] thereby significantly enhancing the non-covalent interactions between molecules. This results in a substantial upward shift in frequencies of acoustic bands and of optical rigid-body bands. The oscillations associated with the low-lying internal branches (5.5–6.4 THz at 0 GPa) are still governed by weak intermolecular interactions, as they primarily comprise distortions of the entire molecular backbones. Therefore, they are also significantly influenced by the molecular packing, as bending and torsion modes affecting the entire molecule become energetically more costly under compression due to a tighter packing. Their pressure-induced upward shifts in frequency are, however, still smaller than for the external vibrations. Consequently, the aforementioned band gap shrinks under increasing pressure and vanishes when the pressure exceeds 2.1 GPa. This can be seen in the band structure Figures 5b and c and becomes obvious from the evolution of the densities of states (DOSs) in Figure 5d. In fact, at higher pressures, the states become more uniformly distributed across the entire displayed spectral range (up to 10 THz; 333 $cm^{-1}$). Since the total number of states remains constant, the growing spacing between phonon bands goes hand in hand with a reduction of the (average) magnitude of the DOSs.

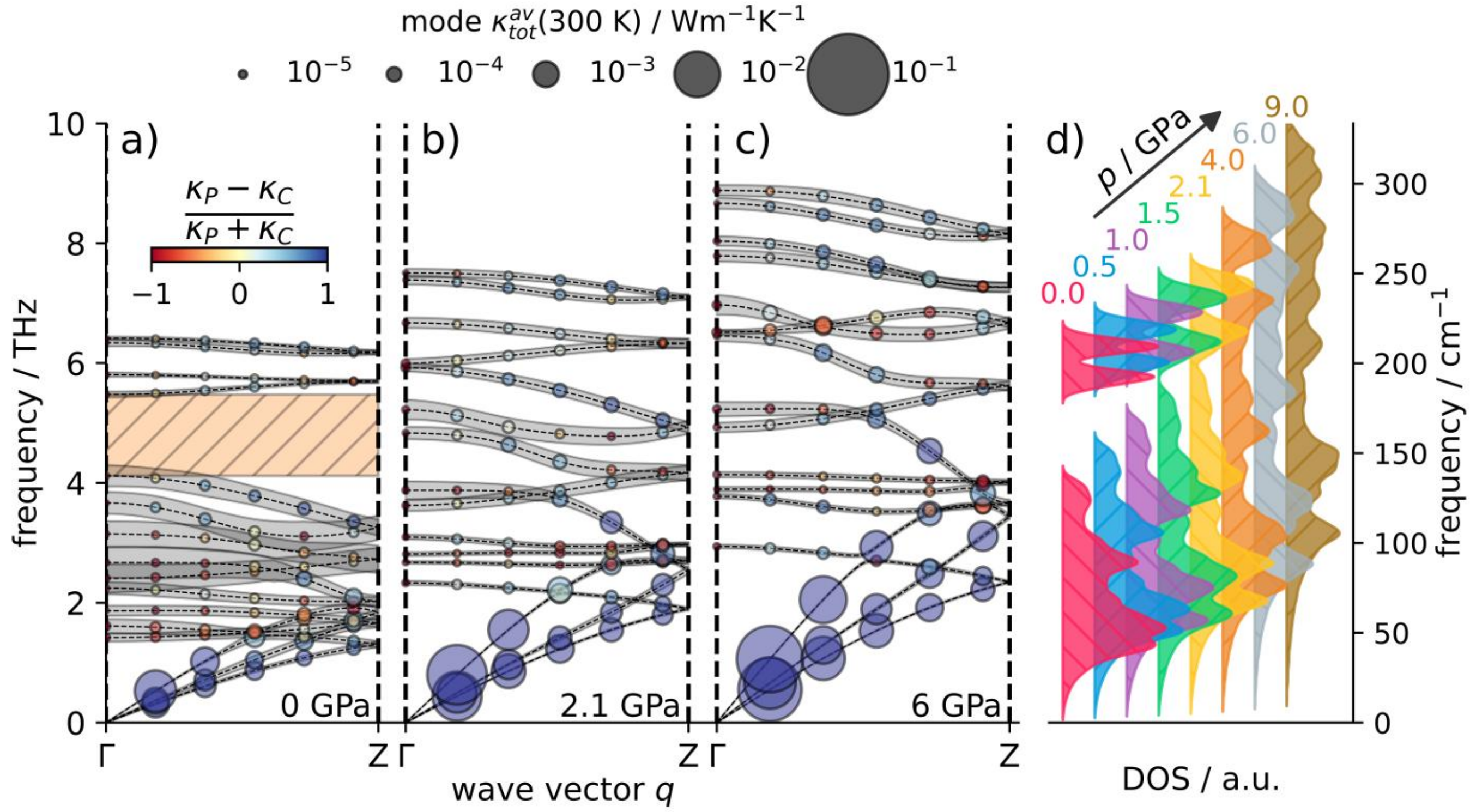

**Figure 5**: *Low-frequency phonon band structure and mode contributions to the total orientation-averaged thermal conductivity. (a-c) Dispersions along high-symmetry path Γ-Z with q= (0,ξ,0) for naphthalene at 0 GPa, 2.1 GPa, and 6 GPa. Anharmonic phonon linewidths (Lorentzian FWHM of spectral function) at 300 K are shown by the grey shading. Mode contributions to $\kappa_{tot}^{av}$ are represented by the diameters of the colored circles, with colors indicating the dominant conduction mechanism (propagation or tunneling). The respective color bar is contained in panel a. (d) Phonon densities of states (DOSs) at select pressures between 0 and 9 GPa, calculated on a 16×22×16 q-mesh with a broadening of 0.075 THz.*

To correlate phonon bands with their thermal conductivity contributions, Figure 5a-c shows the mode-specific contributions of phonons along the high-symmetry paths in reciprocal space. The magnitude of the respective contribution is represented by the size of colored data points (see legend at the top of the panels). The color of each data point indicates the dominant transport mechanism: blue for propagation-dominated modes and red for tunneling-dominated modes (see color bar in Figure 5a). To visualize the anharmonicity present at 300 K phonon linewidths (Lorentzian FWHM of the spectral function), $\Gamma(\mathbf{q})$, are shown as gray shading of each phonon band. They are derived from the three-phonon scattering rates applied here in the evaluation of the Wigner transport equation.

As an additional piece of information, Figure 6a presents the spectral thermal conductivity, $\kappa(\omega)$, as a function of frequency for pressures ranging from 0 to 6 GPa. The data in Figures 5 and Figure 6a corroborate earlier statements in three key aspects: (i) elevated pressures redistribute the conducting phonons across a broader frequency range. This results in an upward shift of the maximum of $\kappa(\omega)$. (ii) The pressure-induced enhancement of the thermal conductivity, observed also in experiments (cf. Figure 2), primarily stems from a growing propagation conductivity (cf., Figure 3a). (iii) and that most of it can be traced back to acoustic bands (cf., Figure 4a), where transverse phonons exhibit a stronger response than longitudinal ones (cf., Figure 4b). Additionally, the band structures in Figure 5 show that the linewidths of external vibrations below 4 THz narrow noticeably under pressure, while the four internal bands (in Figure 5a above 5.5 THz) exhibit a broadening.

To more explicitly illustrate trends for the phonon properties entering the Wigner transport equation, Figures 6b and 6c show the phonon modes' (normed) group velocities and lifetimes. They are sampled on the 8×11×8 $q$-meshes used for calculating the thermal conductivities at various pressures. The data for modes with predominantly acoustic character (APR ≥ 0.5) are denoted by filled symbols, while the properties of modes with mostly optical-character are plotted as open symbols.

Figure 6b explicitly shows what can already be inferred from the band structures, namely the that the pressure-induced frequency upshifts enhance band dispersion and lead to increased group velocities. The effect is strongest for the acoustic phonons. As phonon group velocities contribute quadratically to $\kappa_P$ (Equation 2), this is one of the key factors for the strongly enhanced mode contributions to the thermal conductivity, especially in the region of the acoustic phonons (c.f., Figure 5a-d and Figure 6a). Interestingly, also for several of the optical bands the group velocities increase with pressure, albeit to a lesser degree. On a semi-quantitative level, Figure 6b indicate that from 0 to 6 GPa the group velocities nearly double. This should result in roughly a fourfold increase in the thermal conductivity from 0 to 6 GPa. The actual increase of $\kappa_P$ is, however, ten-fold (see Figure 3a). Hence, the rising group velocities alone are insufficient to fully account for the observed effect.

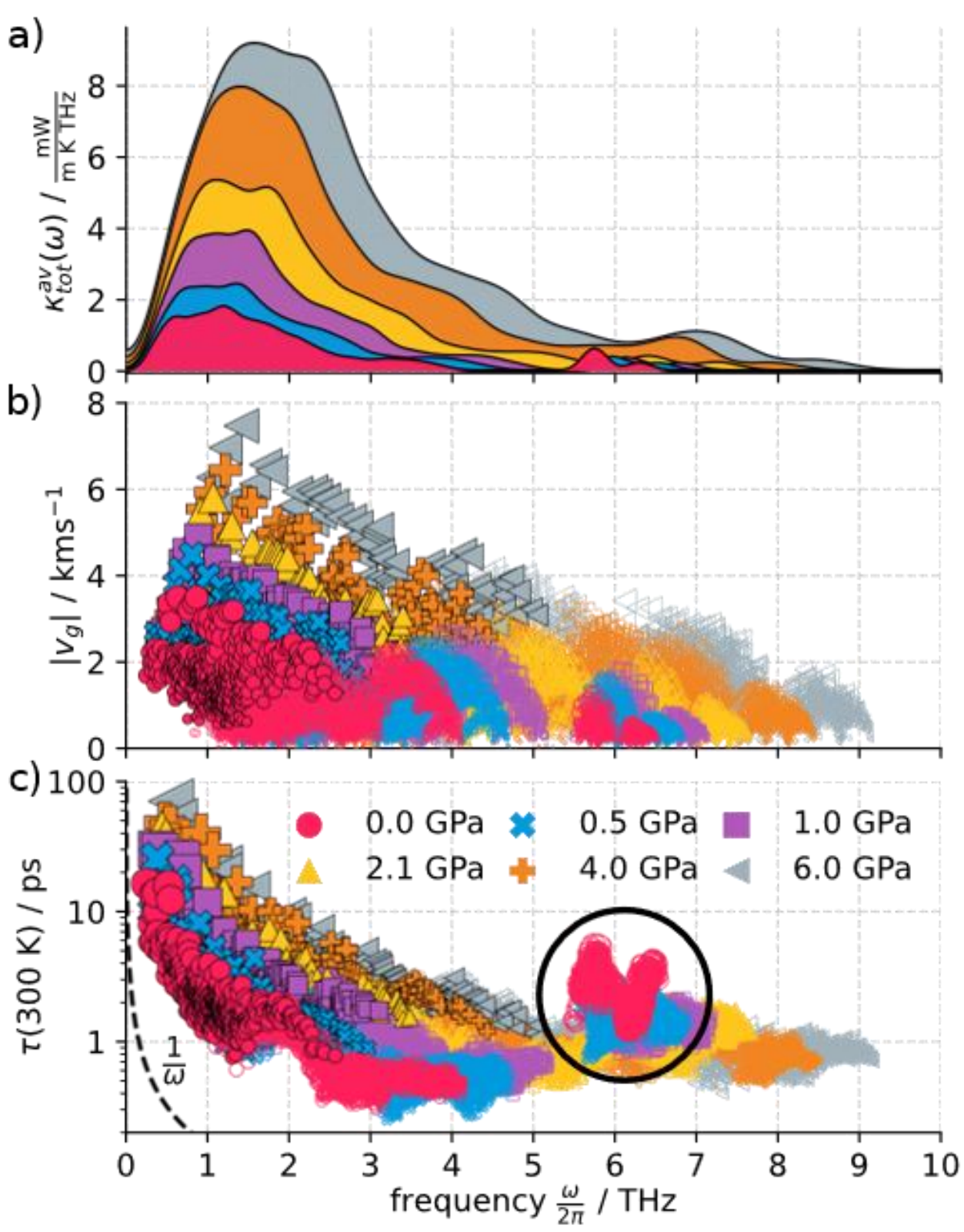

**Figure 6:** *Spectral thermal conductivity and phonon properties of naphthalene at various pressures as a function of vibrational frequency: (a) spectral total orientation-averaged thermal conductivity at 300 K for pressures between 0.0 GPa and 6.0 GPa. A Gaussian broadening, increasing with pressure, between 0.11 and 0.29 THz has been applied. (b) shows the norm of group velocities, and (c) the lifetimes at 300 K (note the log. scale). The Ioffe-Regel limit[70] $\omega^{-1}$ is indicated by a dashed, black line. Symbols and colors denoting different pressures are listed in the legend in panel c. Filled (open) symbols denote modes with predominantly acoustic (optical) character.*

As $\kappa_P$ additionally grows linearly with phonon lifetime, $\tau$, it is useful to consider the pressure-enhancement of the lifetimes. These are shown in Figure 6c. High pressures raise the lifetimes of the low-frequency phonons, as spreading phonon states over a larger energy range diminishes the phase-space for scattering processes. The reduction in allowed scattering channels results in an up-to fourfold increase of the acoustic phonons' lifetimes, further boosting the thermal conductivity.[3] A more in-depth analysis reveals that lifetimes of TA modes increase more strongly than those of the LA modes. This explains why TA phonons become the leading heat carriers at elevated pressures (c.f., Figure 4b). The pressure-induced reduction in scattering is also evident for the other nine intermolecular optical bands (below ~4 THz at 0 GPa). They exhibit progressively narrower linewidths under pressure (gray shadings, Figures 5a-c). In contrast, for the backbone-distorting vibrations, which form the four optical bands (between 5.5 THz and 10 THz at 0 GPa), lifetimes decrease with pressure (see black circle in Figure 6c). This results in a broadening of the respective bands with pressure (Figures 5a-c). This observation can be attributed to the pressure-induced closure of the band gap between inter- and intramolecular phonons, which opens up new three-phonon scattering channels. While the neglect of higher-order scattering could overestimate phonon lifetimes near the ~6 THz gap at low pressures (an artifact discussed, for example, in Ref. 54), we expect this effect to be minor. The excellent agreement with experimental data supports this view, and any such artifact would be diminished at high pressure when the band gap closes.

## Conclusion

In the present contribution, we demonstrate that machine-learned potentials can bridge the gap between first-principles accuracy and the needed computational efficiency to predict the thermomechanical properties of molecular organic semiconductors even under high pressures. This is achieved by including high-pressure reference structures in the training approach of the used MLPs. It allows capturing the non-linear and anisotropic compression of naphthalene with an accuracy comparable to DFT (Figure 1 b,d). It also enables the prediction of phonon scattering rates in naphthalene across a wide range of pressures, a task intractable for DFT. In this way, the trained multi-pressure MTP provides a number of unprecedented and reliable insights into the mechanisms of heat conduction in a soft, anharmonic, and crystalline material under extreme conditions.

A first relevant aspect is the type of heat-transport mechanism. In previous studies it has been argued that with increasing anharmonicity and size of the unit cell, a crystal becomes more complex, which would induce non-negligible heat transport via mode-coupling similar to what occurs in amorphous solids.[43,44,71,72] We have observed this previously when studying the increasing complexity of systematically longer molecules in the series of acene crystals.[42] For naphthalene under high external pressure, one observes the opposite. At ambient pressure, naphthalene indeed behaves like an anharmonic, complex crystal. Densely spaced bands together with strong scattering and hence, broad phonon linewidths, facilitated by the bands' close proximity, lead to a large number of overlapping phonon states. This in turn results in strong inter-band couplings and a significant tunneling (coherence) contribution that at 300 K makes up 25% of the total thermal conductivity.

However, under increasing external pressure, band dispersions increase (see Figure 5a-d), as the compression of the crystal effectively strengthens the non-covalent bonds between the naphthalene molecules. This boosts the phonon group velocities and phonon lifetimes (as key parameters for classical, particle-like phonon transport; c.f., Figure 6b and c). As a consequence, heat conduction via the particle-like propagation mechanism becomes increasingly dominant. This, together with the underlying reduction in phonon scattering can be interpreted as a decrease of the material's anharmonicity. In that spirit, naphthalene transforms from a "complex material" where the tunneling conductivity is significant into a "simpler" crystal with a diminished tunneling contribution when high hydrostatic pressure is applied (<5% at 300 K and 9 GPa). Hence, one can observe a crossover from the Wigner to the Boltzmann transport regime.

Beyond the transport mechanism, pressure also redefines the roles of optical and acoustic phonons as heat carriers. The dominance of optical phonons at ambient pressure gives way to a regime where acoustic phonons become the primary conductors of heat at high pressures via the particle-like mechanism (Figure 4b). The emergence of acoustic phonons as primary heat carriers is another characteristic of a simple crystal, reinforcing the notion of a transition towards a more conventional transport behavior with pressure.

Coinciding with the crossover in dominant transport mechanism and heat carrier type, the thermal conductivity undergoes a significant enhancement that rivals most hitherto studied materials[1]. Still, despite a four-fold increase at 2.1 GPa (the highest measured pressure in [27]), naphthalene's conductivity reaches only 1.53 $Wm^{-1}K^{-1}$ and at best comes into the range of typical thermoelectric materials such as $Bi_2Te_3$ and PbTe.[73,74] Only at pressures exceeding 9 GPa does naphthalene's thermal conductivity surpass 5 $Wm^{-1}K^{-1}$, reaching values comparable to common crystalline minerals such as quartz ($SiO_2$) or rutile (TiO).[75,76] While still modest by inorganic standards, such a conductivity is quite significant for an organic crystal. It is particularly striking given that non-covalent interactions, such as van der Waals forces, between the molecular building blocks are highly detrimental to thermal transport. The fact that hydrostatic pressure can strengthen these non-covalent interactions enough for naphthalene to approach conductivities comparable to those in covalently bonded crystals is rather amazing. From a design perspective, this means that reinforcing intermolecular bonds in molecular crystals, for example, through more polarizable terminal substituents, could lead to thermal conductivities far surpassing those hitherto observed in organic semiconductors.

# Methods

For this article, we applied a workflow to train system-specific machine learned potentials (MLPs) for efficient and accurate simulations of complex organic crystal, similar to what is described in [48,49,42]. The workflow is based on active-learning[46,47] molecular dynamics (MD) simulations within the Vienna Ab-initio Simulation Package (VASP; version 6.4.1)[77–79]. These are performed to generate high-quality datasets computed with dispersion-corrected density functional theory (DFT) for the purpose of training MLPs. Before discussing the active-learning protocols, we first describe the underlying DFT methodology. We then outline the MLP training procedure, followed by the lattice dynamics methods used to compute pressure-dependent thermal conductivity.

**Details of the applied density functional theory method**

Density functional theory calculations, either during on-the-fly active learning or for determining pressure-dependent lattice parameters, were performed combining the Perdew-Burke-Enzerhof (PBE) exchange-correlation functional[58] with the standard projector-augmented wave functions[80] and the dispersion correction by Grimme with Becke-Johnson damping (D3BJ; "IVDW = 12")[59]. For pressure-dependent relaxations of the primitive unit cells, a Γ-centered 2×3×2 *k*-point grid was used. Active-learning simulations were performed for (1,2,1) supercells and the *k*-points were adapted to a 2×2×2 grid. A plane-wave energy cutoff of 1050 eV (800 eV + 30% adjustment for Pulay stress mitigation) was used in both cases. The *k*-point grid and energy cutoff were justified by convergence tests which can be found in the supporting information of [42] (supplementary figure 2). The energy difference for the electronic self-consistency loop was set to $10^{-8}$ eV and a smearing width of 0.05 eV was used for electronic occupations. To generate pressure-dependent configurations during active learning or to determine pressure-dependent lattice parameters by minimizing the enthalpy, an external pressure was applied in VASP. This was achieved by allowing the cell shape to vary ("ISIF = 3") with the "PSTRESS" tag set to a target pressure in kbar.

**On-the-fly active-learning protocols**

In VASP active-learning simulations, the phase space of atomic configurations of a system for target conditions is sampled along an MD trajectory.[46,47,81–83] In the initial phase, a small number of MD steps is calculated with DFT, forming the basis of the reference data. Then, a VASP-internal MLP ("ML-FF") is trained on this data and retrained during the simulation whenever a critical number of new DFT-calculated configurations have been added. This ML-FF is used for the majority of MD steps, while DFT steps are only required when configurations proposed by the ML-FF are too dissimilar from the existing reference data. Whether a DFT step is required is detected by a Bayesian error analysis of atomic forces and a dynamically adjusting error threshold. Thus, the trajectory is corrected, and a reference dataset of DFT-calculated configurations is generated. In this way, the active-learning approach allows efficient sampling of the phase space and ensures that only a manageable number of the "most relevant" configurations are computed with DFT.

For this work, the targeted MLP use case was lattice dynamics simulations requiring accurate forces from small displacements around the equilibrium positions. To generate training data most relevant to this use case, we employed low to intermediate simulation temperatures (50–200 K) to sample phase space near the equilibrium structure, as discussed in detail in [49] and [42]. These MD simulations were performed in an *NpT* ensemble using a Langevin thermostat with friction coefficients of 10 $ps^{-1}$ for each atom type and the lattice degree of freedom. For the latter, we also assigned a fictitious mass of 1000 amu. Otherwise, default values for the Bayesian error threshold, the ML-FF descriptors and other machine-learning hyperparameters of VASP version 6.4.1. were used.

Building upon the basic training set (previously established in [42]), each subsequent active-learning simulation under external pressure incorporated previously collected training data to prevent redundant DFT calculations of similar configurations. All simulations were performed on a (1×2×1) supercell of naphthalene (72 atoms). Starting structure for both, the initial 0 GPa and the 0.5 GPa simulations, were based on the same DFT-relaxed unit cell. For the following simulations at higher pressures, we continued from the final output structure of the preceding MD run to maintain the progressive cell compression. The basic training dataset at 0 GPa comprised 459 configurations collected along a 15,000-step trajectory with a 0.5 fs time interval. During the three consecutive 5,000-step simulations, at 0.5 GPa, 2 GPa, and 10 GPa the data set was subsequently expanded with 222, 163, and 168 configurations, respectively.

**Multi-pressure moment tensor potential**

Following the procedure detailed in [42], three level 28 moment tensor potentials (MTPs) with a radial basis size of 20, a radial cutoff of 5 Å, and a minimum distance between atoms of 0.95 Å were independently trained on the multi-pressure training set using the MLIP-2 package[61,62]. The MTP with the smallest root-mean-square deviation (RMSD) of phonon frequencies from the DFT reference[45,84] at 0 GPa was retained for calculating further properties. The basic MTP achieved a frequency RMSD of 1.8 $cm^{-1}$ for low frequencies and 2.4 $cm^{-1}$ over the entire spectrum. The multi-pressure MTP produces comparable RMSDs of 2.1 $cm^{-1}$ for low-frequency modes and 3.0 $cm^{-1}$ overall. The essentially identical RMSD at low frequencies is consistent with the above observation that the multi-pressure MTP provides an excellent description of ambient pressure thermal conductivities (c.f. Figure 2).

**Pressure-dependent structure relaxation**

DFT-relaxation of naphthalene's lattice parameters and atomic positions while applying an external pressure was performed with VASP, employing a tolerance for the maximum force component of $10^{-3}$ eV$Å^{-1}$.

Relaxing the lattice and the atomic positions with the multi-pressure MTP was a two-step process: first, lattice and ionic positions are relaxed with a routine in MLIP-2[61,62] with a force tolerance of $10^{-4}$ eV$Å^{-1}$ and a stress tolerance of $10^{-3}$ eV$Å^{-2}$. Then, the ionic positions were additionally relaxed applying a stricter force difference threshold of $10^{-8}$ eV$Å^{-1}$ with a routine utilizing LAMMPS[85] and the LAMMPS-MLIP interface[86].

**Lattice dynamics**

To determine the lattice vibrations, interatomic force constants (FCs) were calculated using a finite-differences scheme on (2,3,2) supercells of the compressed structures (432 atoms), in which one (2$^{nd}$ order FCs) or two (3$^{rd}$ order FCs) atoms were displaced. For the 2$^{nd}$ order FCs, the validity of this supercell dimensions was tested for both 0 GPa and 9 GPa (see Supporting Information; Figure S4). The displaced structures were generated with either Phonopy or Phono3py.[87,88] As an outcome of convergence tests, displacement amplitudes of 0.01 Å were chosen for 2$^{nd}$ order FCs (see Supporting Information; Figure S2) and 0.05 Å for 3$^{rd}$ order FCs (see Supporting Information; Figure S5) for all pressure values. No cut-off limitations for 3$^{rd}$ order FCs had to be imposed due to the use of an efficient force calculator. Atomic forces of displaced supercell were then again calculated with a python script that employed LAMMPS and the LAMMPS-MLIP interface. In a subsequent step, the interatomic FCs were obtained with either Phonopy or Phono3py accounting for translational and index-exchange symmetries.

Phonons and their scattering rates, velocity operators, and mode heat capacities were evaluated on a mesh of 8×11×8 q-points in the primitive unit cell using Phono3py. Phonon scattering rates were calculated with the tetrahedron method for Brillouin zone integration.[89] The corresponding lattice thermal conductivities are calculated with the Wigner transport equation[43,44] within the single-mode relaxation time approximation (RTA) as implemented in Phono3py. The validity of using the RTA compared to the direct solution of the linearized Boltzmann transport equation (LBTE)[90] has been explicitly tested for naphthalene and this test can be found in the supporting information of [42] (supplementary figure 7). For the thermal conductivity convergence with regards to the supercell dimensions of 3$^{rd}$ order FCs and q-point meshes, we also refer to the supporting information of [42] (supplementary figure 9 and supplementary figure 10). Additionally, convergence with regard to the q-mesh for compressed naphthalene was explicitly tested for a pressure of 2.1 GPa, which can be seen in Figure S6 of the Supporting Information.

## Acknowledgements

This research was funded in whole, or in part, by the Austrian Science Fund (FWF) [primarily Grant-DOI: 10.55776/P33903 and in part also 10.55776/P36129]. Computational results have been obtained using the Austrian Scientific Cluster. M.S. acknowledges the ACES resource at Texas A&M University through allocation TRA120004 from the Advanced Cyberinfrastructure Coordination Ecosystem: Services & Support (ACCESS) program, which is supported by U.S. National Science Foundation grants #2138259, #2138286, #2138307, #2137603, and #2138296.

The authors acknowledge the use of Academic AI to improve the grammar, spelling, and clarity of language.

We note that the recent crystal structure investigation of naphthalene up to 50 GPa by Zhou et al.[91] came to our attention only after the present manuscript was largely completed and could therefore not be considered in our analysis and discussion.

## Data Availability Statement

All data supporting the findings of this study will be made publicly available upon publication of the manuscript.

## Code Availability Statement

VASP can be acquired from the VASP Software GmbH (see www.vasp.at/); LAMMPS is available at www.lammps.org; MLIP-2 is available at gitlab.com/ashapeev/mlip-2; the LAMMPS-MLIP interface is available at gitlab.com/ashapeev/interface-lammps-mlip-2; Phonopy is available at phonopy.github.io/phonopy; Phono3py is available at phonopy.github.io/phono3py. Python scripts for the data analysis will be made publicly available with the respective research data.

# Supporting Information

# Origins of Pressure-Enhanced Thermal Transport in Organic Semiconductors

Lukas Legenstein[1,2,*], Sandro Wieser[3], Michele Simoncelli[4], Egbert Zojer[2,†]

[1]Chair of Physics, Montanuniversität Leoben, Leoben, 8700
[2]Institute of Solid State Physics, Graz University of Technology, Graz, 8010
[3]Institute of Materials Chemistry, TU Wien, Wien, 1060
[4]Department of Applied Physics and Applied Mathematics, Columbia University, New York, USA

[*]lukas.legenstein@unileoben.ac.at, [†]egbert.zojer@tugraz.at

## 1. Testing moment tensor potentials against density functional theory

Validation and selection of the multi-pressure moment tensor potentials (MTPs) for naphthalene was performed by comparing phonon frequencies predicted by three independently parameterized MTPs against those calculated using the same density functional theory (DFT) method employed in training. The root-mean-square deviations (RMSDs) between DFT-calculated and MTP-predicted phonon band frequencies are presented in Table S1 for three distinct spectral ranges. The multi-pressure MTP used throughout this work was selected based on the lowest frequency RMSDs ("MTP#1" in Table S1). For visual comparison, Figure S1 shows an overlay of the DFT phonon band structure with that of "MTP#1", which exhibited the highest accuracy. The DFT band structure stems from ref. 45 and the respective data was obtained from [92] (10.17172/NOMAD/2021.09.28-1). Band structures from DFT (blue) and the multi-pressure MTP (orange) are mostly identical, with only miniscule frequency discrepancies.

***Table S1.*** *Root-mean-square deviations (RMSDs) between DFT-calculated phonon frequencies and those predicted by three independently parameterized moment tensor potentials (MTPs) along high-symmetry paths in the reciprocal space of naphthalene. RMSDs are reported for three frequency ranges: (i) below 5 THz (including all intermolecular vibrations), (ii) below 20 THz (most relevant for thermal transport), and (iii) below 100 THz (including all phonon branches). The RMSDs are provided in units of terahertz and wavenumbers (in brackets).*

| | RMSD($f$<5 THz) | RMSD($f$<20 THz) | RMSD($f$<100 THz) |
|---|---|---|---|
| MTP #1 | **0.061 THz (2.0 cm $^{-1}$)** | **0.086 THz (2.9 cm $^{-1}$)** | **0.090 THz (3.0 cm $^{-1}$)** |
| MTP #2 | 0.205 THz (6.8 cm $^{-1}$) | 0.180 THz (6.0 cm $^{-1}$) | 0.161 THz (5.4 cm $^{-1}$) |
| MTP #3 | 0.067 THz (2.2 cm $^{-1}$) | 0.112 THz (3.7 cm $^{-1}$) | 0.106 THz (3.5 cm $^{-1}$) |

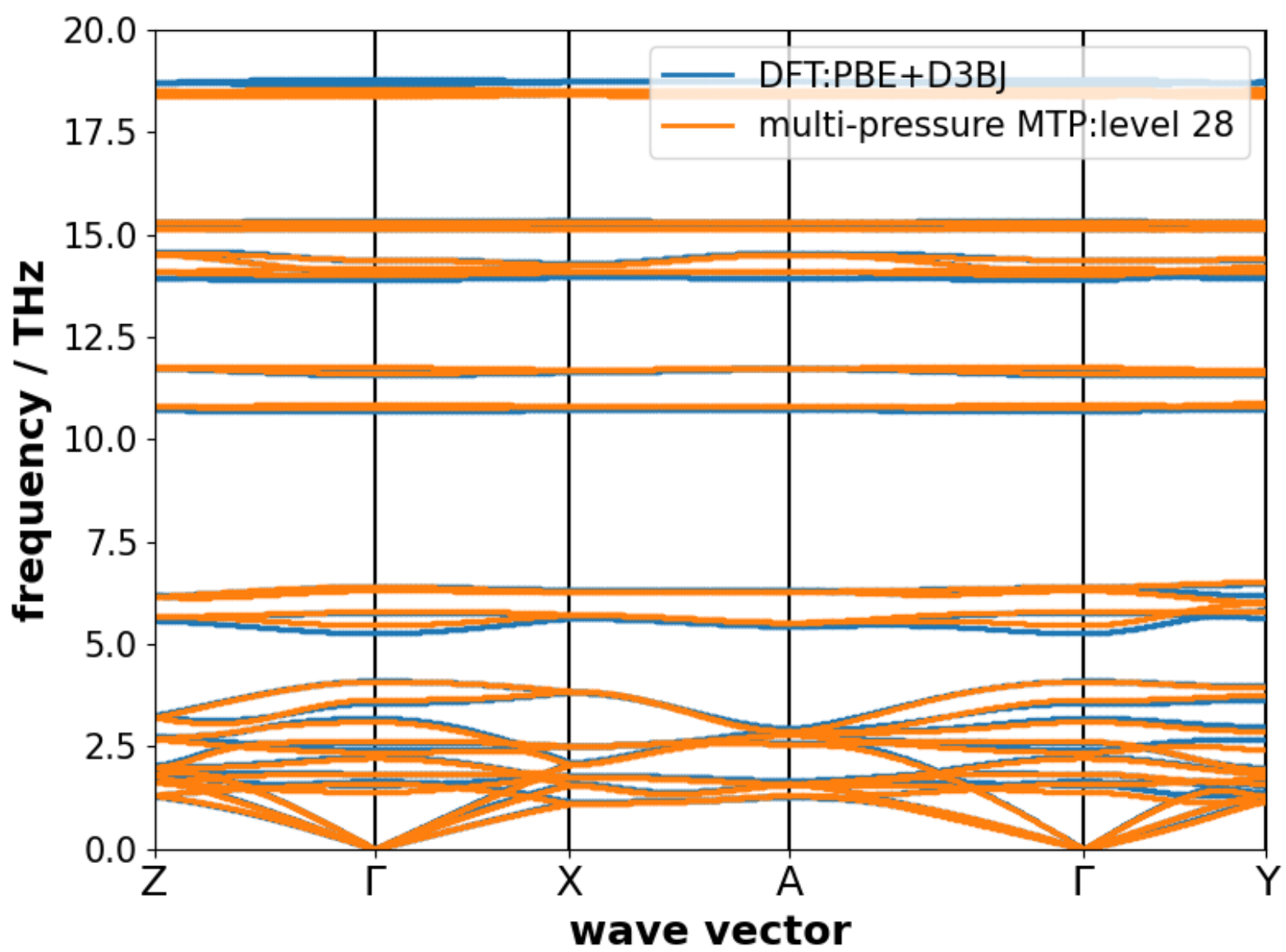

***Figure S1:*** *Phonon band structure of naphthalene up to 20 THz, calculated with atomic forces from density functional theory (blue) and from the chosen multi-pressure moment tensor potential (orange).*

To validate the structural predictions of our multi-pressure moment tensor potential (MTP), we compared its pressure-dependent lattice constants with those obtained from DFT relaxations. We performed DFT relaxations for pressures between 0 and 16 GPa using VASP and MTP relaxations for pressures between 0 and 19 GPa (methodological details in the main text Methods section). Figure S2 presents these results alongside experimental reference data which are room-temperature XRD measurements from Likhacheva et al. and neutron diffraction data at 5 K from Capelli and colleagues.[30,32]

The MTP shows excellent agreement with DFT predictions up to 15 GPa, after which the MTP results begin to deviate. We observe a minor discrepancy in the triclinic angle $\beta$ between 2.5 and 5 GPa. As demonstrated by the close agreement between the cryogenic neutron diffraction experiments and our computational results, the offset between the 295 K XRD data and our theoretical predictions results from thermal expansion which is discussed below.

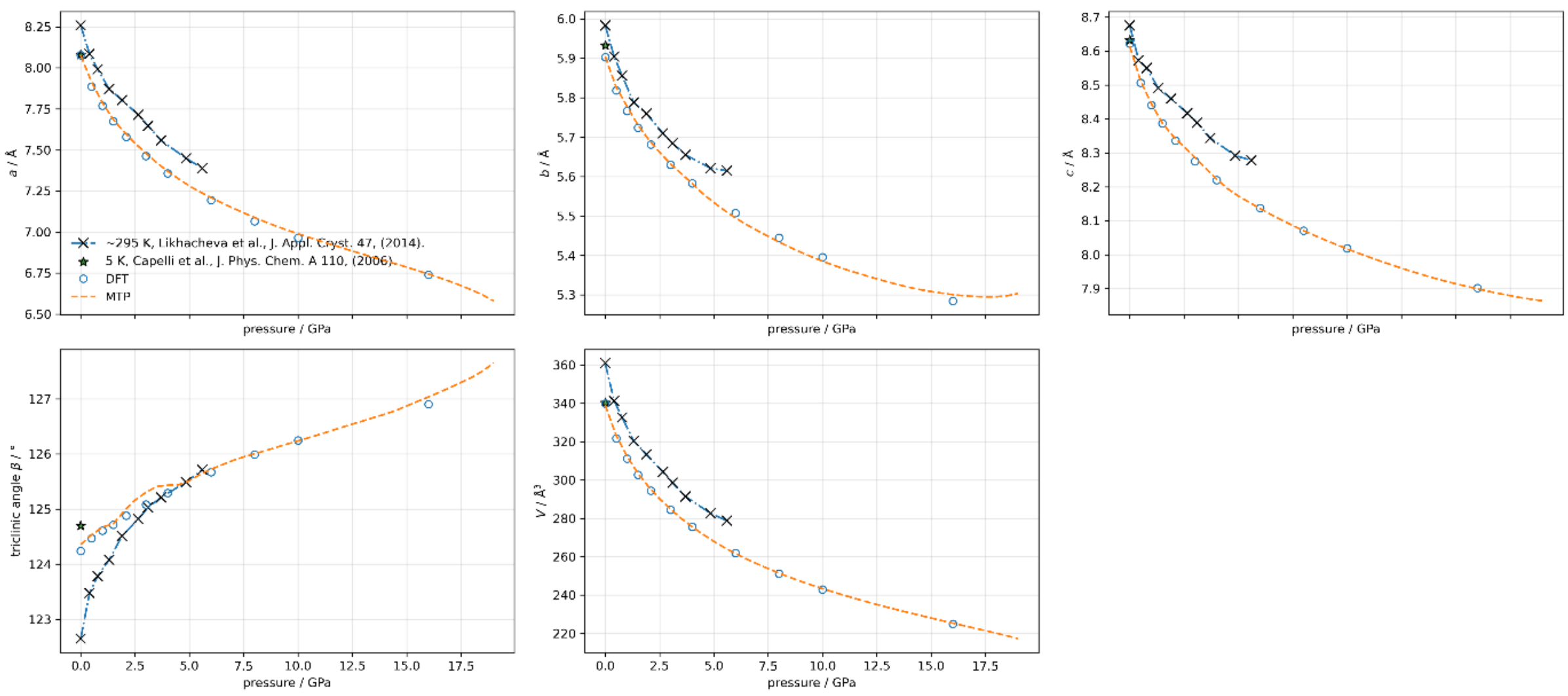

***Figure S2:*** *Pressure-dependent lattice parameters (a, b, c, and β) and unit cell volume (V) of naphthalene. The multi-pressure MTP predictions are shown as dashed orange lines, DFT-calculated values as blue circles, room-temperature experimental data from Likhacheva et al.*[32] *as black crosses, and the 5 K neutron diffraction parameters from Capelli et al.*[30] *at 0 GPa as a black star.*

## 2. Impact of thermal expansion

In the main work, compressed crystal structures of naphthalene were obtained through gradient-based enthalpy optimization, as implemented in the MTP-native MLIP-2 package[61] or VASP[77–79]. This approach does not account for temperature dependence and thus neglects thermal expansion, which can be significant in soft molecular crystals like naphthalene (cf. Figure S2). Here, we argue that this approximation has negligible impact on our thermal conductivity predictions, particularly at elevated pressures.

As an initial reference point, an explicit test in the supporting information of ref. 42 (Supplementary Figure 21) demonstrated that including thermal expansion reduces naphthalene's thermal conductivity at 300 K by only ~8%. To further assess the role of thermal expansion in our pressure-dependent predictions, Figure S3 compares the orientation-averaged thermal conductivity from the MTPs with experimental data by Ross et al.[27] (green crosses) up to 0.75 GPa. However, their data does not include measurements at ambient pressure because their measurement technique required a minimum pre-compaction pressure to ensure adequate contact of the measurement wire with the naphthalene powder. We therefore performed a linear fit to extrapolate their data to 0 GPa.

At ambient pressure, the basic MTP prediction exhibits a slight upward deviation from the otherwise linear pressure trend. This deviation is consistent with the neglect of thermal expansion. When we instead calculate the thermal conductivity using the experimental 295 K unit cell from Capelli et al.[35], the prediction coincides with the extrapolated experimental trend. At pressures above ~0.1 GPa, however, the agreement between our enthalpy-optimized cell predictions and the experimental data is excellent without any such correction.

These findings are consistent with prior literature observations. Nicol and colleagues[93] reported that Raman-active vibrational frequencies of naphthalene become nearly temperature-independent above 0.5 GPa. Similarly, Likhacheva et al. observed a "drastic decrease of thermal expansion of naphthalene [...] between zero and 1 GPa" and noted "In the case of naphthalene, we observe nearly zero thermal expansion at 18% volume contraction (≈3 GPa) ..."[33] attributing this behavior to a general feature of herringbone-packed molecular crystals. Together with our own findings, these observations strongly support the conclusion that thermal expansion effects are negligible for the pressure range considered in the main work.

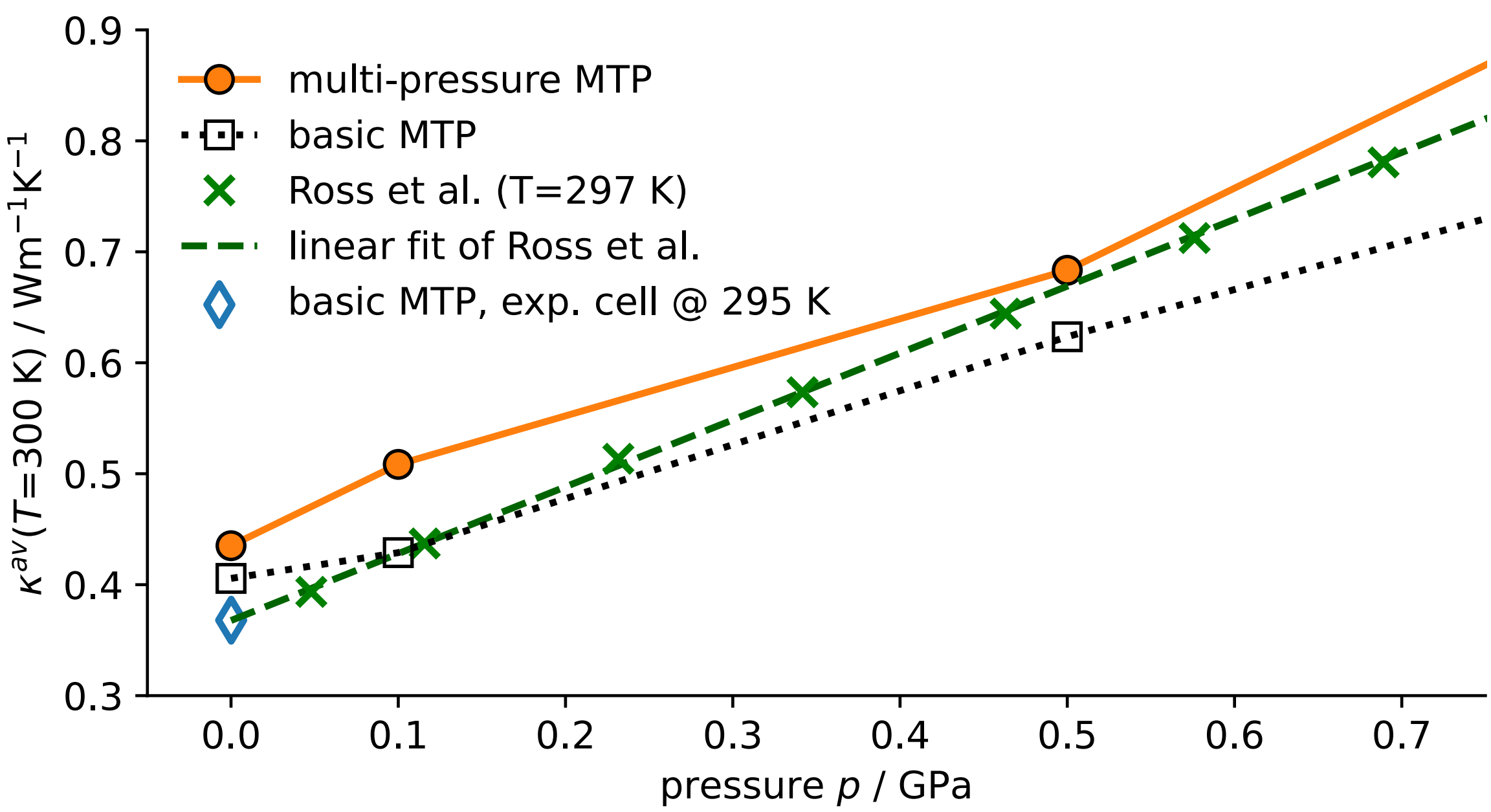

***Figure S3:*** *Orientation-averaged thermal conductivity of naphthalene as a function of pressure in the range up to 0.75 GPa. Included are data from the multi-pressure MTP (orange circles), the basic MTP (open squares), and experiments by Ross et al.[27] (green crosses). A linear fit through the latter is denoted by a dashed green line. A separate basic MTP prediction for the neutron-diffraction-determined unit cell at 295 K and ambient pressure (Capelli et al.[30]) is presented by a blue, open square.*

### 3. Convergence of harmonic phonon frequencies

We verified the convergence of phonon frequencies with respect to both displacement amplitude and supercell dimensions (Figure S4). Convergence tests were performed for structures at 0 GPa and 9 GPa external pressure. Based on these tests, we selected a (2,3,2) supercell and 0.01 Å displacement amplitude for all calculations across the entire pressure range.

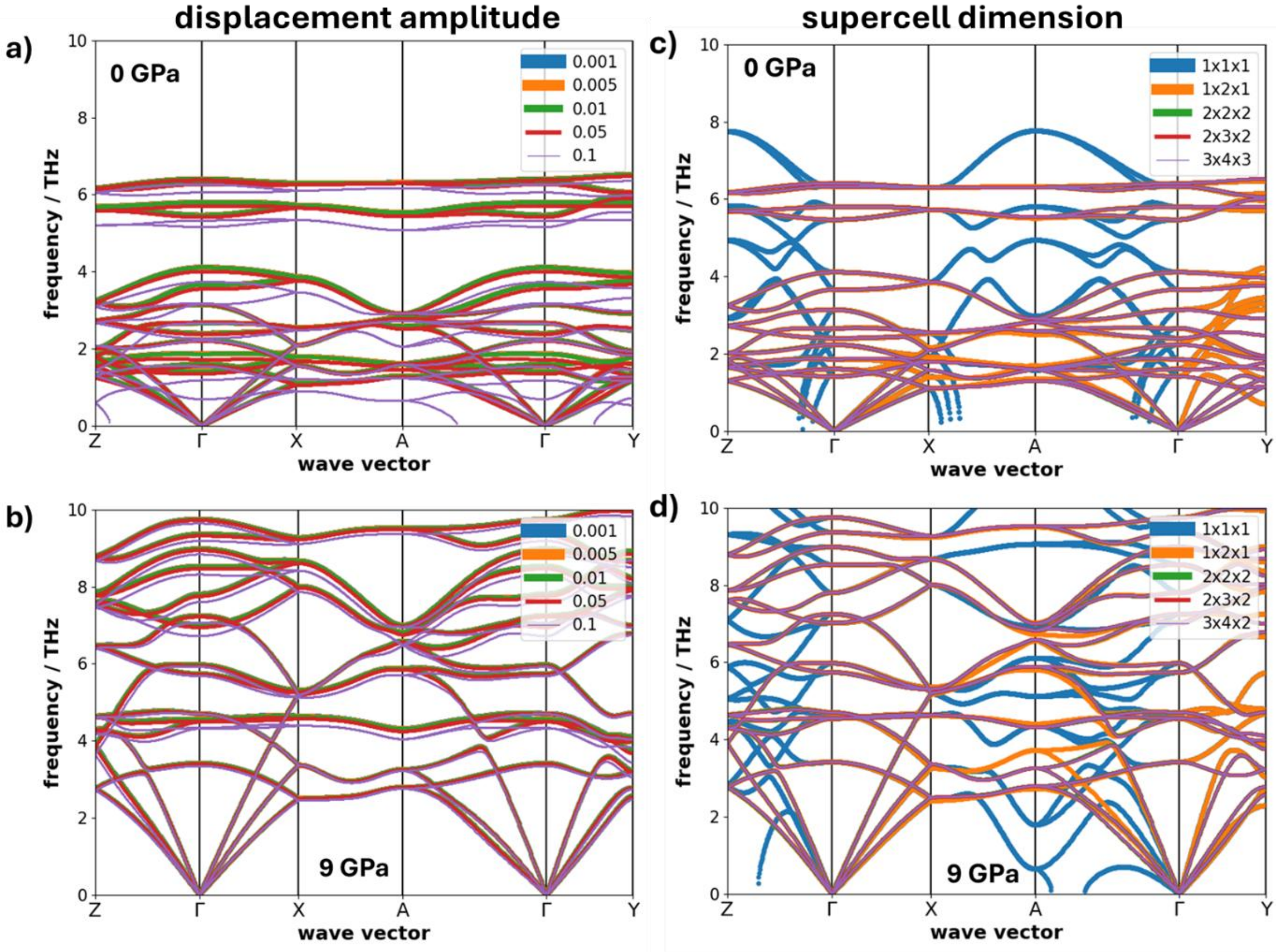

***Figure S4:*** *Convergence tests for phonon band frequencies w.r.t. the finite differences displacement amplitude (panels a and b) and w.r.t. the supercell dimension (panels c and d), both at 0 GPa and 9 GPa. Data for different displacement amplitude or supercell dimensions are presented with different colors according to the panels' respective legends.*

## 4. Convergence of lattice thermal conductivities

Several parameters influence results from Peierls-Boltzmann or Wigner transport equation simulations. While these parameters have been tested at ambient pressure in the supporting information of reference [42], we now test whether the displacement distance and size of the q-point mesh need to be adapted for elevated pressures.

Rather than testing parameters rigorously for each pressure, we sought a common displacement distance suitable across all conditions. We calculated the orientation- and temperature-averaged thermal conductivity of naphthalene using several displacement distances at different pressures for the converged 8×11×8 q-mesh (Figure S5). From that, we are finding that 0.05 Å provides consistent results for 0 GPa, 2.1 GPa, 6 GPa, and 8 GPa, covering the pressure range discussed in the main work. In addition to the calculations with the multi-pressure MTP, Figure S5 includes thermal conductivities at 0 GPa from the basic MTP, for which displacement distances from 0.03 Å to 0.07 Å produce converged values.

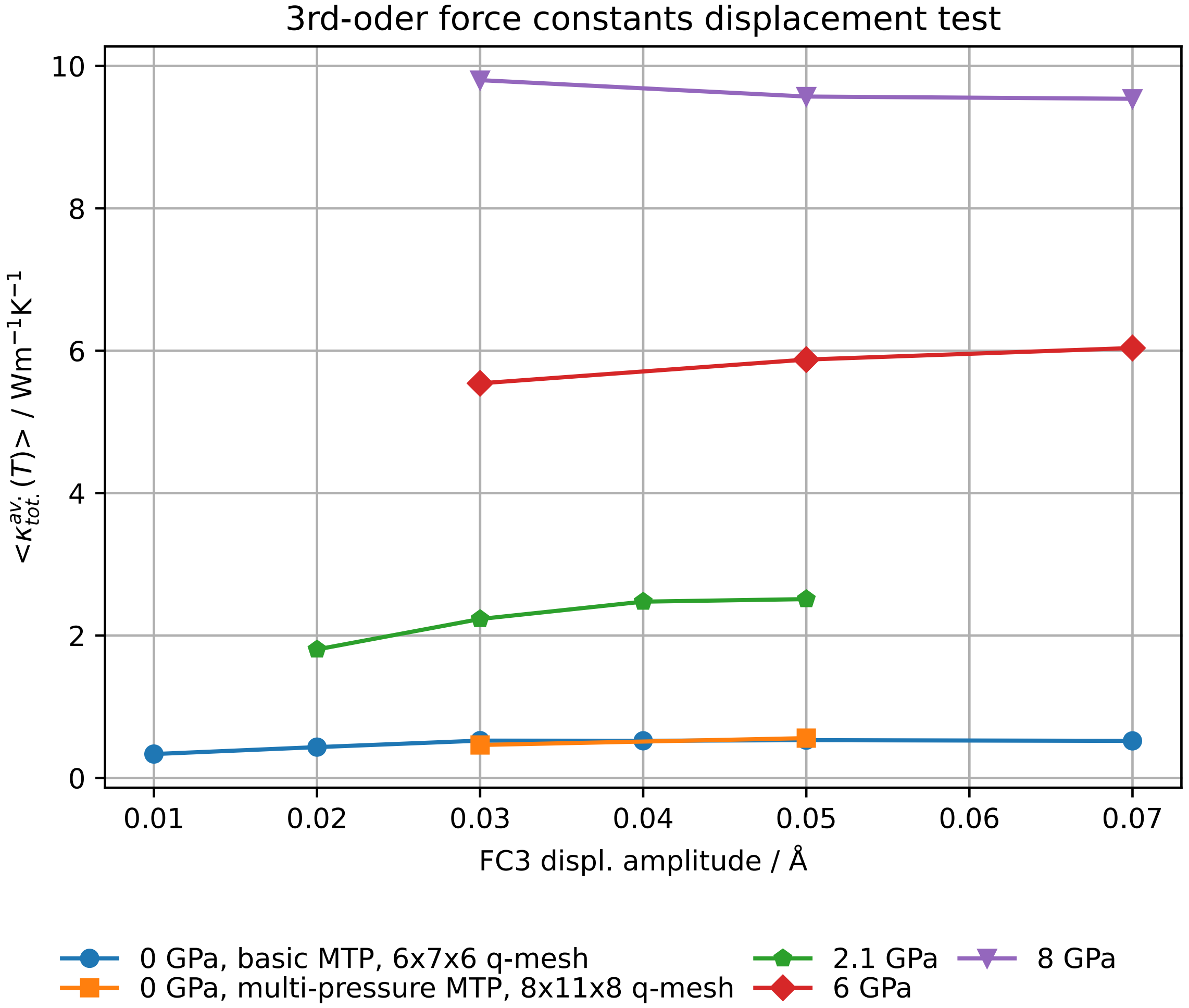

***Figure S5:*** *Orientation- and temperature-averaged lattice thermal conductivity of naphthalene as a function of the finite differences displacement amplitude for obtaining the anharmonic 3$^{rd}$ order force constants. Data for different external pressures are represented by the marker symbols and colors according to the legend below the graph.*

In addition to displacement amplitude, we examined whether the q-mesh size requires adjustment with increasing pressure. This parameter may be critical because hydrostatic pressure reduces the unit cell volume, potentially necessitating a higher q-point density. Furthermore, the pressure-enhanced group velocities might require finer sampling. Based on the aforementioned convergence tests at ambient pressure from earlier work (which included q-mesh evaluation), we initially assumed an 8×11×8 mesh would be sufficient. To verify this, we compared this mesh against a significantly denser 11×15×11 mesh by calculating the orientation-averaged thermal conductivities at 2.1 GPa (Figure S6). The results show negligible differences across the entire studied temperature range.

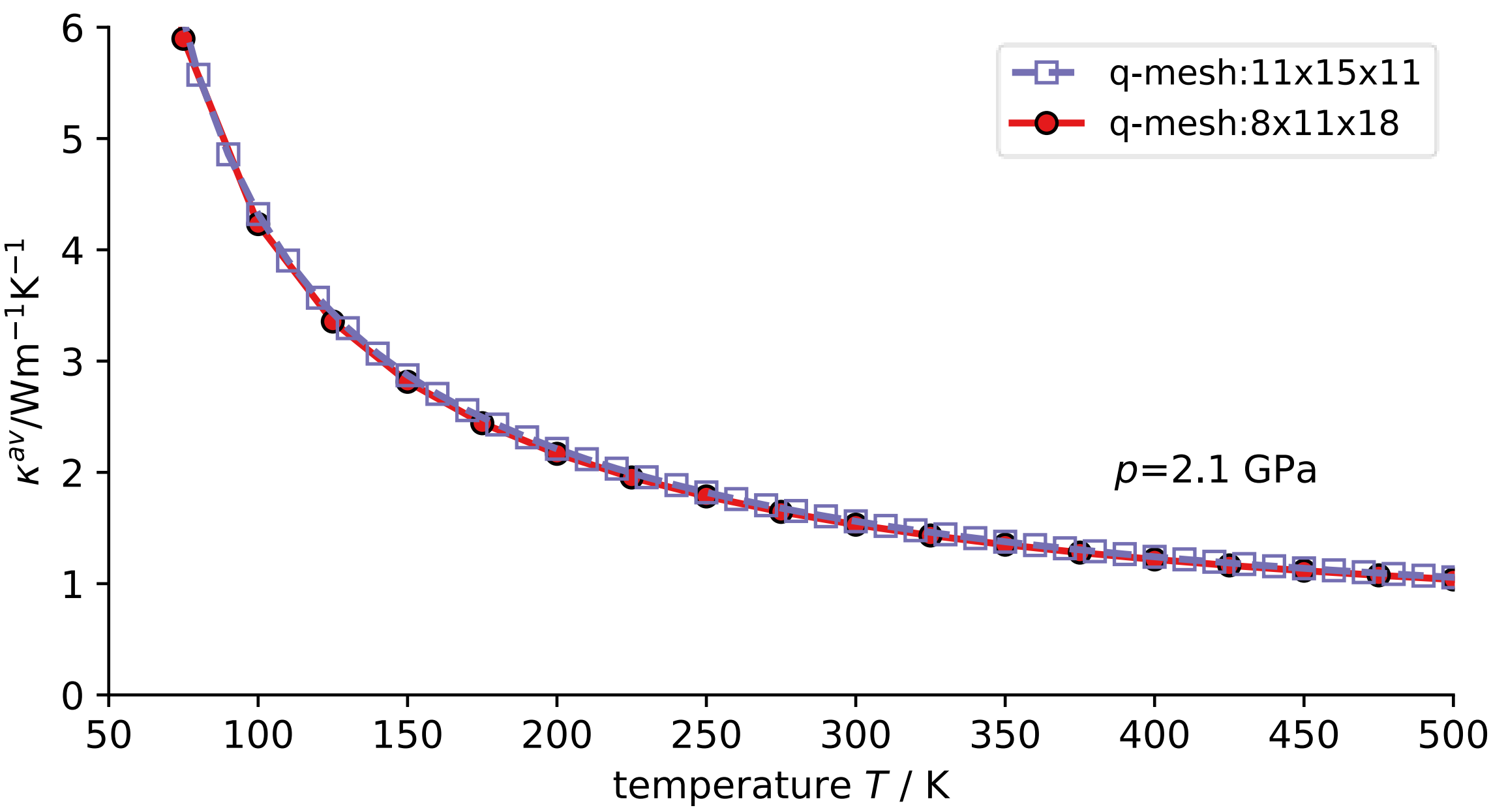

***Figure S6:*** *Orientation-averaged, total thermal conductivity of naphthalene as a function of pressure from solving the Wigner transport equation for q-point meshes with dimensions 8×11×8 (red, filled circles) and 11×15×11 (purple, open squares).*

**5. More details on pressure-dependent thermal conductivity**

The pressure-dependent thermal conductivities shown in the main work (Figure 2) were limited to the experimentally accessible range. In Figure S7, we extend these calculations to 15 GPa. The linear trend observed at lower pressures continues at least up to this pressure. Beyond 15 GPa, the reliability of our MTP's structure predictions decreases compared to density functional theory results (see Figure S2).

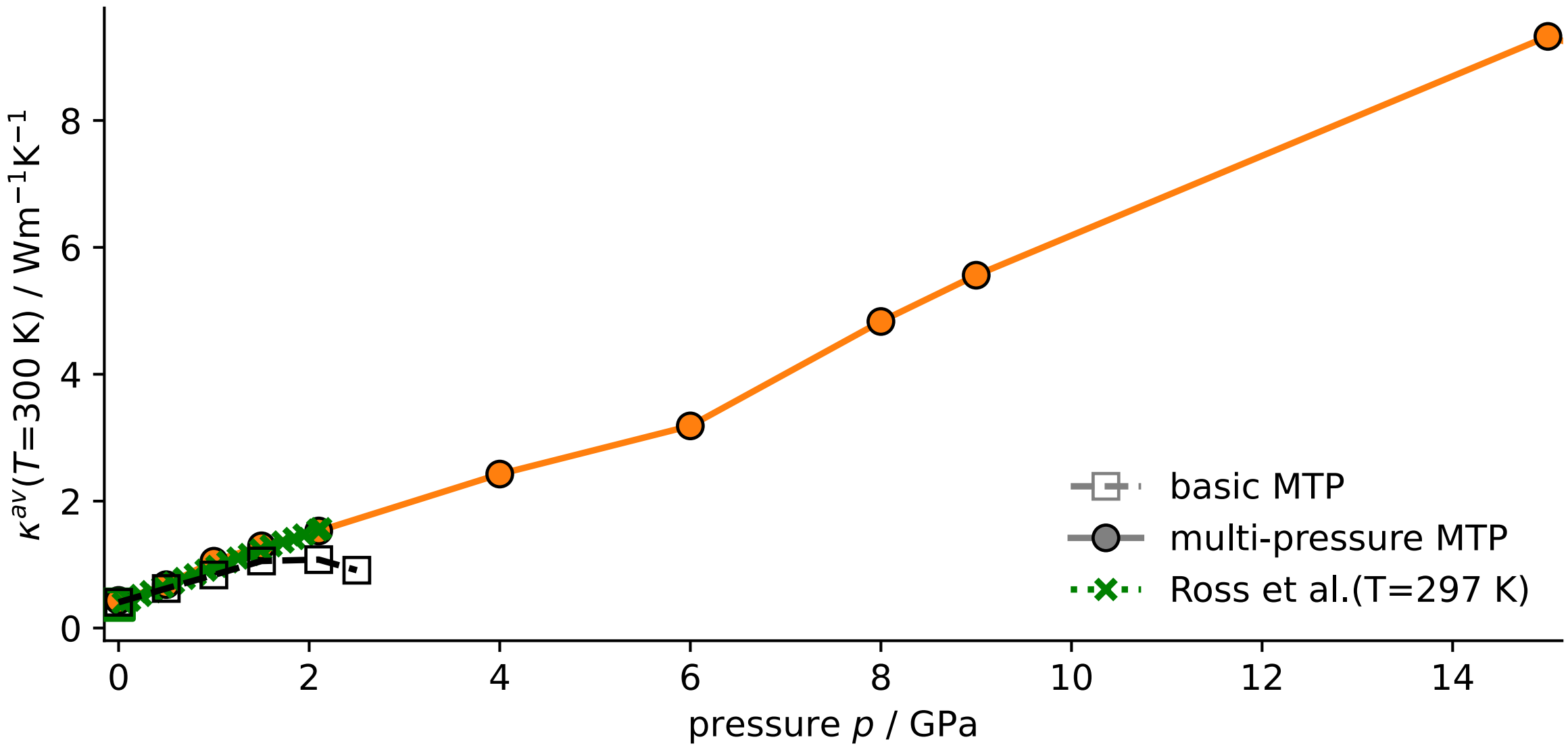

***Figure S7:*** *Orientation-averaged, total thermal conductivity of naphthalene as a function of pressure at room temperature: from the basic MTP (open squares), from the multi-pressure MTP (orange circles), from experiments by Ross et al.*[27] *(green crosses).*

### 6. Details on decomposed acoustic/optical thermal conductivity contributions

A visual representation of the spatial alignment between the phonon's displacement vectors and their wave vectors ("longitudinality") is provided in Figure S8 by showing naphthalene's band structure at 9 GPa. This property provides a more robust method for identifying longitudinal acoustic phonons than simply selecting the third-lowest frequency band. In the figure, bands colored in red indicate modes where atomic displacements are nearly parallel to the wave vector, characteristic of longitudinal acoustic phonons. This approach allows us to track longitudinal character even through (avoided) band crossings or mode hybridization events.

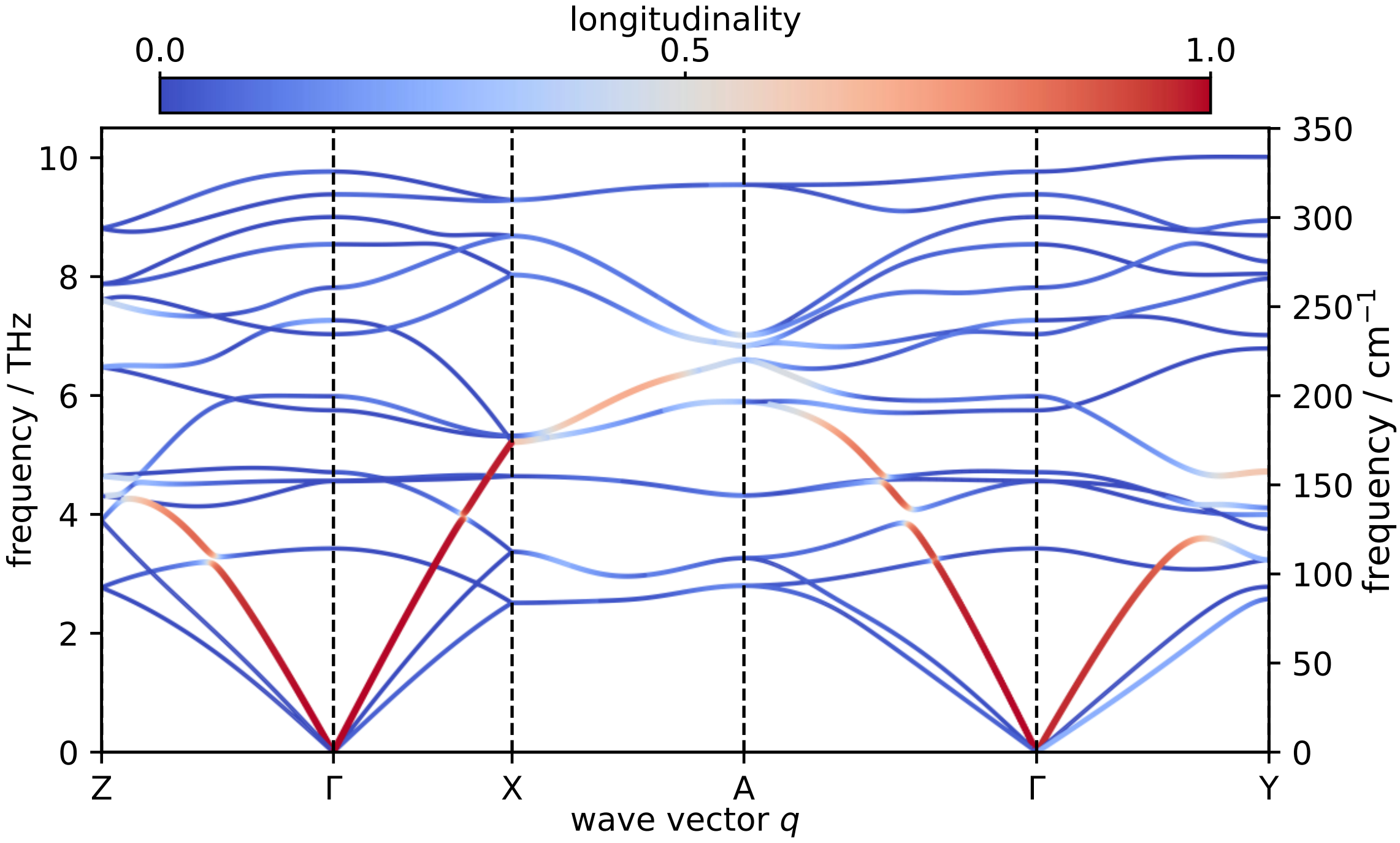

***Figure S8:*** *Phonon band structure of naphthalene at 9 GPa of hydrostatic pressure. The color code indicates the "longitudinality" of the individual phonon modes (see color bar), where those in red can be characterized as longitudinal phonons while the blue ones are not. Intermediate colors indicate modes of mixed character.*

In the main text (Figure 4b), we presented the pressure-dependent relative contributions of longitudinal acoustic, transverse acoustic, and optical phonons to the intraband propagation thermal conductivity ($\kappa_P$). Here, we extend this analysis to the interband tunneling (coherence) conductivity ($\kappa_C$). For this mechanism, we decompose the contributions based on coupling between pairs of phonon bands rather than individual phonon modes.

Figure S9a shows the absolute pressure-dependent contributions from acoustic-acoustic (AA), acoustic-optical (AO), and optical-optical (OO) coupling, while Figure S9b presents their relative contributions. Our results reveal that AA coupling contributes negligibly to thermal conductivity. AO coupling accounts for 10-30% of $\kappa_C$ (depending on temperature), with this contribution decreasing as pressure increases. The majority of the tunneling conductivity arises from OO coupling.

For completeness, Figure S10 shows the absolute contributions of longitudinal acoustic, transverse acoustic, and optical phonons to the propagation thermal conductivity ($\kappa_P$) as a function of pressure, complementing the relative contributions shown in Figure 4b.

.

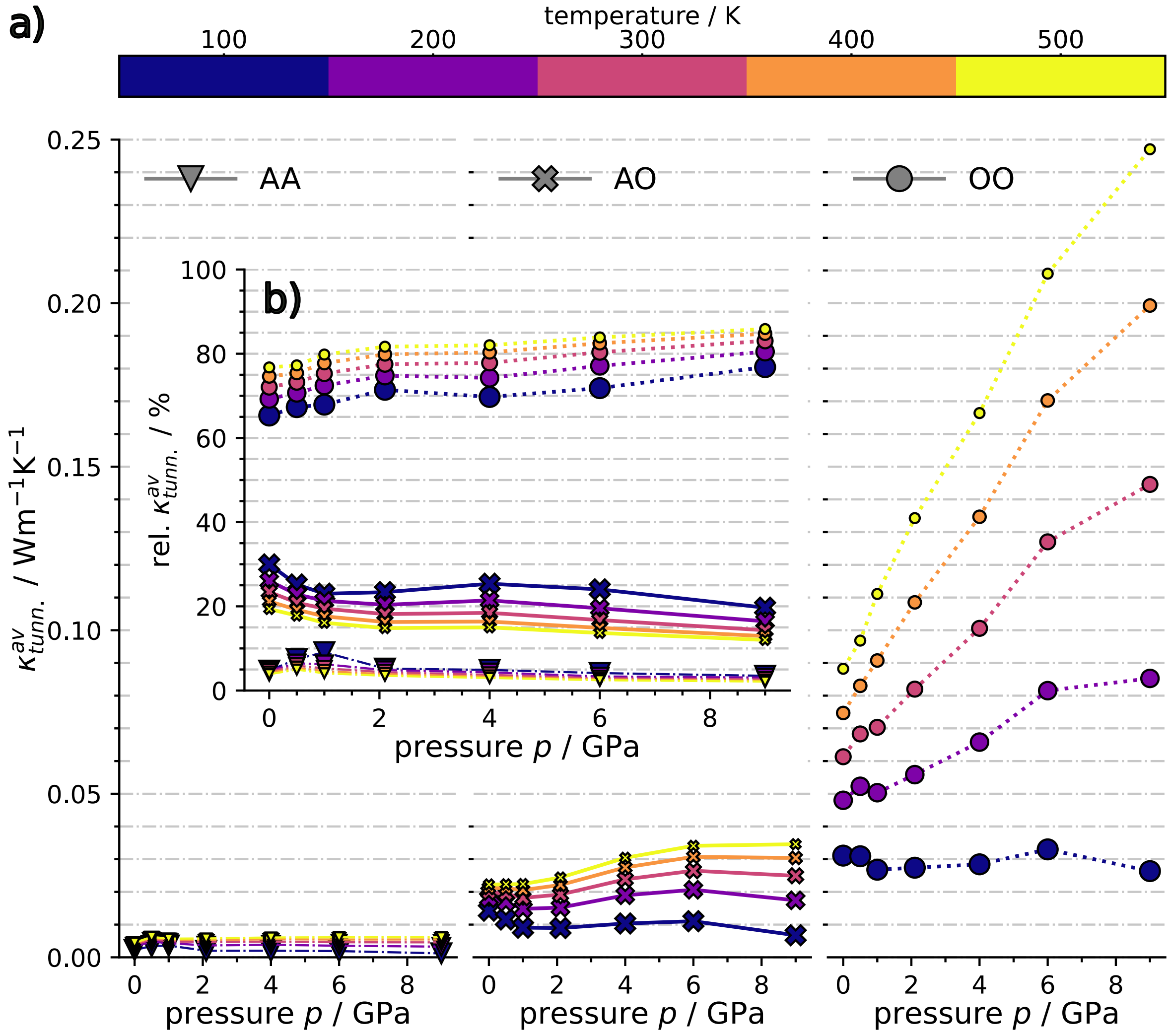

**Figure S9:** *Decomposition of the orientation-averaged tunneling (coherence) thermal conductivity as a function of pressure: interband coupling between two acoustic bands (AA) is denoted by triangles, between an acoustic and an optical band (AO) by crosses, and between two optical bands (OO) by circles. The color of the data points denotes the temperature according to the color bar at the top of panel a. Panel a shows the absolute values of the tunneling (coherence) thermal conductivity and panel b shows the relative values of the decomposed contributions to* $\kappa_C$.

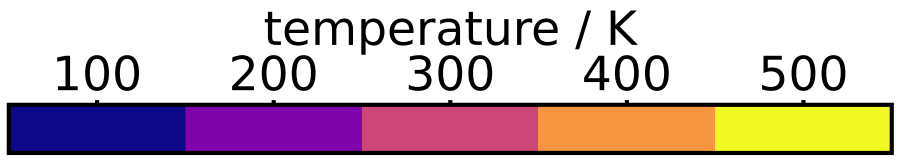

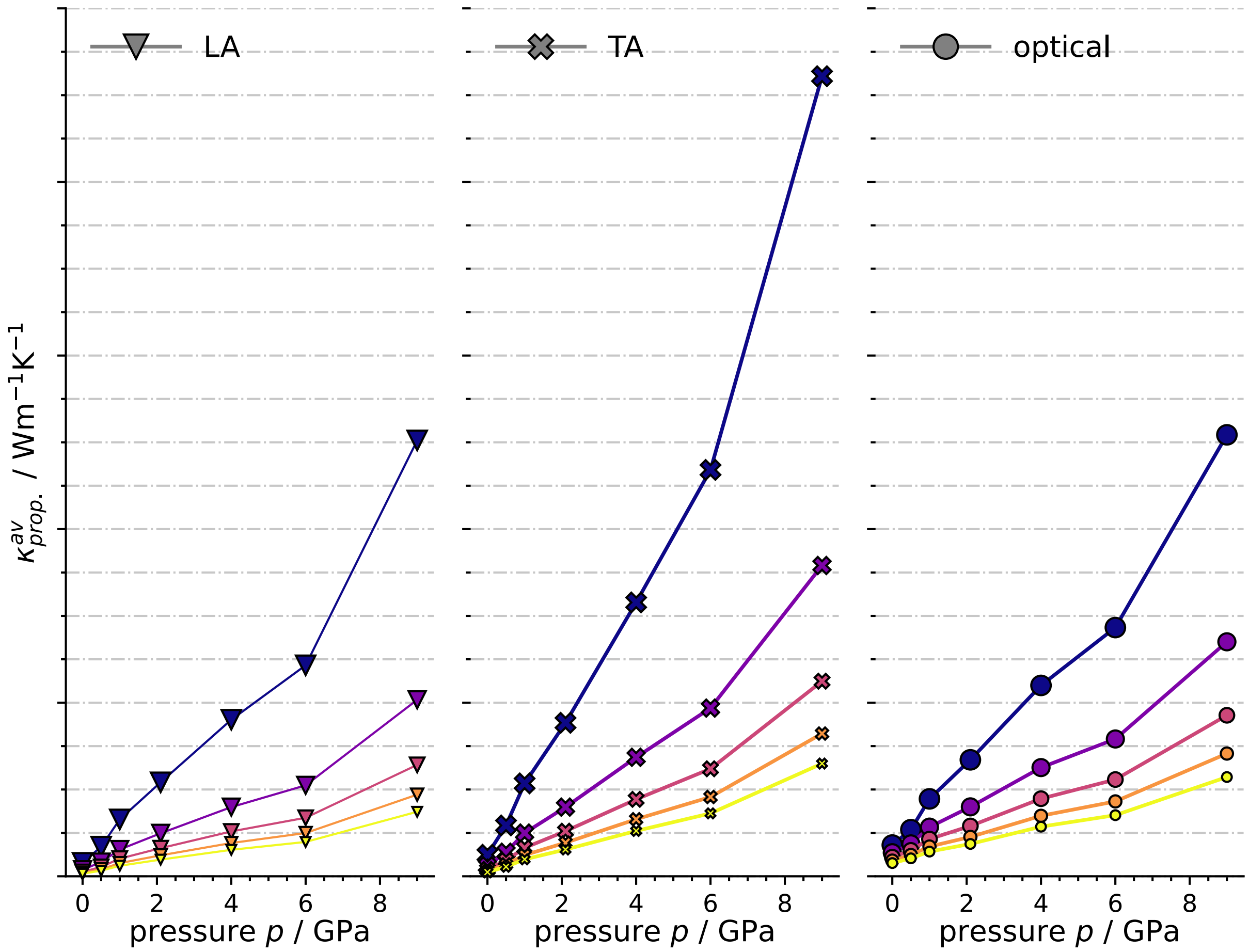

**Figure S10:** *Absolute contributions of optical (circles), transverse acoustic (crosses), and longitudinal acoustic (triangles) phonons to the orientation-averaged propagation thermal conductivity as a function of pressure. The color of the data points denotes the temperature according to the color bar at the top of the figure.*